\documentclass[format=acmsmall]{acmart}
\usepackage{booktabs} 
\usepackage[english]{babel}
\usepackage[export]{adjustbox}
\usepackage{mathrsfs}
\usepackage{csquotes}
\usepackage{multirow}
\usepackage{multicol}
\usepackage{array}
\usepackage{arydshln}
\usepackage{booktabs}
\usepackage{subfig}
\usepackage{siunitx}
\usepackage[inline]{enumitem}
\usepackage{xspace}
\usepackage{etoolbox}
\usepackage[skip=2pt]{caption}


 
 \acmJournal{TOIS}
\acmVolume{0}
\acmNumber{0}
\acmArticle{0}
\acmYear{2019}
\acmMonth{0}
\copyrightyear{2019}

\newlist{inlinelist}{enumerate*}{1}
\setlist*[inlinelist,1]{%
  label=(\roman*),
}
\author{Mohammad Aliannejadi}
\authornote{Part of the work reported in this paper was done while Mohammad Aliannejadi was affiliated with the Universit\`a della Svizzera italiana (USI).}
\affiliation{%
  \institution{University of Amsterdam}
  \city{Amsterdam} 
  \country{The Netherlands}
}
\email{m.aliannejadi@uva.nl}

\author{Hamed Zamani}
\affiliation{%
  \institution{University of Massachusetts Amherst}
  \city{Amherst} 
  \state{MA}
  \country{USA}
}
\email{zamani@cs.umass.edu}

\author{Fabio Crestani}
\affiliation{%
  \institution{Universit{\`a} della Svizzera italiana (USI)}
  \city{Lugano} 
  \country{Switzerland}
}
\email{fabio.crestani@usi.ch}

\author{W. Bruce Croft}
\affiliation{%
  \institution{University of Massachusetts Amherst}
  \city{Amherst} 
  \state{MA}
  \country{USA}
}
\email{croft@cs.umass.edu}

\newcommand{\revision}[1]{#1}

\newcommand{\partitle}[1]{\vspace{2mm}\noindent\textbf{#1}}

\newcommand{\istas}{ISTAS\xspace}
\newcommand{\lsapp}{LSApp\xspace}
\newcommand{\app}{uSearch}
\newcommand{\ntar}{NTAS}
\newcommand{\appname}[1]{\texttt{#1}}
\newcommand{\modelname}{CNTAS\xspace}
\newcommand{\modelnameneusa}{NeuSA\xspace}

\begin{document}

\title{Context-Aware Target Apps Selection and Recommendation for Enhancing Personal Mobile Assistants}

\begin{abstract}

Users install many apps on their smartphones, raising issues related to information overload for users and resource management for devices. Moreover, the recent increase in the use of personal assistants has made mobile devices even more pervasive in users' lives. This paper addresses two research problems that are vital for developing effective personal mobile assistants: \textit{target apps selection} and \textit{recommendation}. The former is the key component of a unified mobile search system: a system that addresses the users' information needs for all the apps installed on their devices with a unified mode of access. The latter, instead, predicts the next apps that the users would want to launch. Here we focus on context-aware models to leverage the rich contextual information available to mobile devices. We design an in situ study to collect thousands of mobile queries enriched with mobile sensor data (now publicly available for research purposes). With the aid of this dataset, we study the user behavior in the context of these tasks and propose a family of context-aware neural models that take into account the sequential, temporal, and personal behavior of users. We study several state-of-the-art models and show that the proposed models significantly outperform the baselines.

\end{abstract}

\thanks{This work was supported in part by the RelMobIR project of the Swiss National Science Foundation (SNSF), and in part by the Center for Intelligent Information Retrieval.
}


\maketitle

\section{Introduction}
\label{sec:intro}

In recent years, the number of available apps on the mobile app market has been growing due to high demand from users, leading to over 3.5 million apps on Google Play Store, for example.\footnote{\url{https://www.thinkwithgoogle.com/advertising-channels/apps/app-marketing-trends-mobile-landscape/}}
As a consequence, users now spend an average of over five hours a day using their smartphones, accessing a variety of applications.\footnote{\url{http://flurrymobile.tumblr.com/post/157921590345/us-consumers-time-spent-on-mobile-crosses-5}} An average user, for example, installs over 96 different apps on their smartphones~\cite{DBLP:conf/wsdm/Baeza-YatesJSH15}.
In addition, the emergence of intelligent assistants, such as Google Assistant, Microsoft Cortana, and Apple Siri, has made mobile devices even more pervasive. These assistants aim to enhance the capability and productivity of users by answering questions, performing actions in mobile apps, and improving the user experience while interacting with their mobile devices. 
Another goal is to provide users with a universal voice-based search interface; however, they still have a long way to go to provide a unified interface with the wide variety of apps installed on users' mobile phones. 
The diversity of mobile apps makes it challenging to design a unified voice-based interface. However, given that users spend most of their time working within apps (rather than a browser), it is crucial to improve their cross-app information access experience.

In this paper, we aim to address two research problems that are crucial for effective development of a personal mobile assistant: \emph{target apps selection} and \emph{recommendation in mobile devices}. Target apps selection is the key component towards achieving a unified mobile search system -- a system that can address the users' information needs not only from the Web, but also from all the apps installed on their devices. 
We argued the need for a universal mobile search system in our previous work~\cite{AliannejadiSigir18}, where our experiments suggested that the existence of such a system would improve the users' experience. 
Target apps recommendation, instead, predicts the next apps that the users would want to launch and interact with, which is equivalent to target apps selection with no query. 

\begin{figure}[t]
    \centering
    \includegraphics[width=0.75\textwidth]{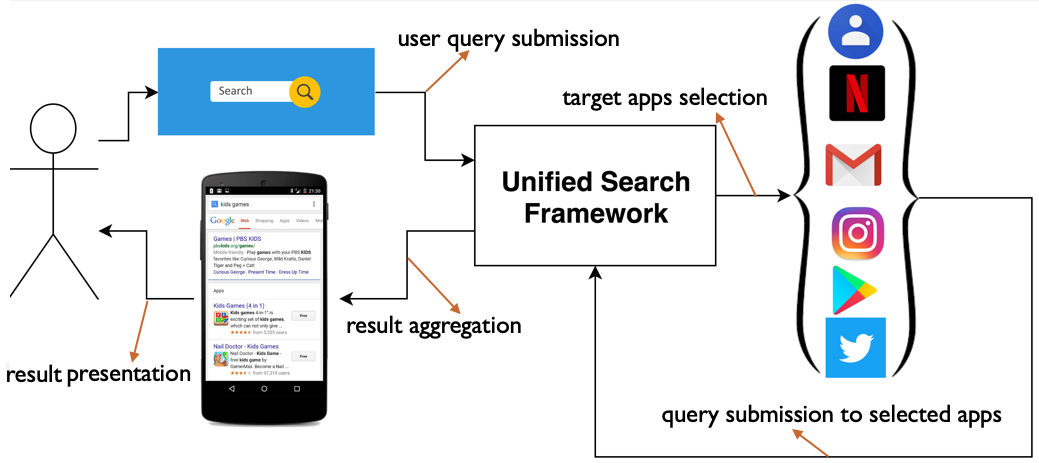}
    \caption{A typical workflow of a unified mobile search framework.}
    \label{fig:ums}
\end{figure}

A unified mobile search framework is depicted in Figure~\ref{fig:ums}. As we see in the figure, with such a framework, the user could submit a query through the system which would then identify the best target app(s) for the issued query. The system then would route the query to the identified target apps and display the results in an integrated interface.
Thus, the first step towards designing a unified mobile search framework is identifying the target apps for a given query, which is the target apps selection task \cite{AliannejadiSigir18}.  

\emph{Target apps recommendation} is also crucial in a mobile environment. It has attracted a great deal of attention in multiple research communities~\cite{DBLP:conf/sigir/BeitzelJCGF04,DBLP:conf/chi/HalveyKS06,DBLP:conf/sigir/ShokouhiG15}. Among various benefits and use cases discussed in the literature, we find the following two cases the most important ones:
\begin{inlinelist}
    \item to assist users in finding the right app for a given task the user wishes to perform;
    \item to help the operating system manage its resources more efficiently.
\end{inlinelist}
It is worth noting that both use cases potentially play essential roles in improving end users' experience. The former reduces the users' effort to find the right app among various apps installed on their phone. On the other hand, the latter can affect the users' experience through smart resource management. For instance, a system could remove many background processes of apps that are not going to be used in the near future to avoid excessive battery usage. It can also be used to allocate the required resources to an app that is going to be launched by the user in the immediate future, providing faster and smoother user experience. The use of a target apps recommendation system and a target apps selection system brings even more benefits. While app usage data can help a target apps selection model provide more accurate predictions, the submitted cross-app queries could also improve a recommendation system's performance. For example, in cases when a user is traveling, they would use travel and navigation apps more often. This could be considered as an indication of the current user's information to the system. Also, assume a user submits the query ``Katy Perry hits'' to \appname{Google}. The recommendation system could use this information in its prediction and recommend music apps. 

As mobile devices provide rich contextual information about users, previous studies \cite{DBLP:conf/mhci/KamvarB07, DBLP:conf/sigir/AliannejadiC17, Zamani:2017:WWW} have tried to incorporate query context in various domains. In particular, query context is often defined as  information provided by previous queries and their corresponding clickthrough data~\cite{DBLP:conf/cikm/WhiteBD10,DBLP:conf/sigir/XiangJPSCL10}, or situational context such as location and time~\cite{DBLP:conf/sigir/BennettRWY11,Zamani:2017:WWW,DBLP:conf/saint/HattoriTT07}. 
However, as user interactions on a mobile device are mostly with apps, exploring apps usage patterns reveals important information about the user contexts, information needs, and behavior. 
For instance, a user who starts spending time on travel-related apps, e.g., TripAdvisor, is likely to be planning a trip in the near future. \citet{DBLP:conf/chi/CarrascalC15} verified this claim by showing that people use certain categories of apps more intensely as they do mobile search.
Modeling the latent relations between apps is of great importance because while people use few apps on a regular basis, they tend to switch between apps multiple times~\cite{DBLP:conf/chi/CarrascalC15}.
In fact, previous studies have tried to address app usage prediction by modeling personal and contextual features~\cite{DBLP:conf/wsdm/Baeza-YatesJSH15}, exploiting context-dependency of app usage patterns~\cite{DBLP:conf/huc/LeeLJJS16}, sequential order of apps~\cite{DBLP:conf/wasa/XuLZGZZZS18} and collaborative models~\cite{DBLP:conf/huc/WangYSZXLC16}. 

However, our previous attempt to study unified mobile search through crowdsourcing did not capture users' contexts in the data collection phase \cite{AliannejadiSigir18} because it was done on the phone's browser, failing to record any contextual and sensor data related to the user location and activities.
In addition, there are some other limitations. For example, we asked workers to complete a set of given search tasks, which obviously were not generated by their actual information needs, and thus the queries were likely different from their real search queries. In addition, not all of  workers completed their tasks on actual mobile devices, which affected their behavior. Furthermore, the user behavior and queries could not be studied over a day-long or week-long continuous period.

These limitations have motivated us to conduct the first in situ study of target apps selection for unified mobile search. This enables us to obtain clearer insights into the task.
In particular, we are interested in studying the users' behavior as they search for real-life information needs using their own mobile devices. Moreover,  we studied the impact of contextual information on the apps they used for search. To this aim, we developed a simple open source app, called \textit{\app}, and used it to build an in situ collection of cross-app queries. 
Over a period of 12 weeks, we collected thousands of queries which enables us to investigate various aspects of user behavior as they search for information in a cross-app search environment. 

Using the collected data, we conducted an extensive data analysis, aiming to understand how users' behavior vary across different apps while they search for their information needs. A key finding of our analysis include the fact that users conduct the majority of their daily search tasks using specific apps, rather than \appname{Google}. Among various available contextual information, we focus on the users' apps usage statistics as their \textit{apps usage context}, leaving others for future work. This is motivated by the results of our analysis in which we show that users often search on the apps that they use more frequently.
Based on the insights we got from our data analysis, we propose a context-aware neural target apps selection model, called \textit{\modelname}. 
In addition, as we aimed to model the sequential app usage patterns while incorporating personal and temporal information, we proposed a neural target apps recommendation model, called \textit{\modelnameneusa}, which is able to predict the next apps that a user would launch at a certain time. 
The model learns complex behavioral patterns of users at different times of day by learning high-dimensional app representations, taking into account the sequence of previously-used apps. 

In summary, the main contributions of this paper are:

\begin{itemize}
    \item An in situ mobile search study for collecting thousands of real-life cross-app queries. We make the app\footnote{\url{https://github.com/aliannejadi/uSearch}}, the  collected search query data\footnote{\url{https://github.com/aliannejadi/istas}}, and the annotated app usage data\footnote{\url{https://github.com/aliannejadi/LSApp}} publicly available for research purposes.
    \item The first in situ analysis of cross-app queries and users' behavior as they search with different apps. More specifically, we study different attributes of cross-app mobile queries with respect to their target apps, sessions, and contexts.
    \item A context-aware neural model for target apps selection.
    \item A personalized sequence-aware neural model for target apps recommendation.
    \item Outperforming baselines for both target apps selection and recommendation tasks.
\end{itemize}

Our analyses and experiments lead to new findings compared to previous studies, opening specific future directions in this research area.

This paper extends our previous work on in situ and context-aware target apps selection for unified mobile search~\cite{DBLP:conf/cikm/AliannejadiZCC18}. We previously stressed the importance of incorporating contextual information in a unified mobile search and studied the app usage statistics data to identify the user's intent of submitting a query more accurately. We showed that considering what applications a person has used mostly in the past 24 hours is useful to improve the effectiveness of target apps selection. In this paper, we further explore the effect of sequential app usage behavior of users for target apps recommendation. This is an ideal complement to our context-aware target apps selection model as these two components constitute an important part of context-aware mobile computing~\cite{DBLP:series/sbcs/CrestaniMS17}. In summary, this paper extends our previous work as follows:
\begin{itemize}
    \item It presents a novel personalized time-aware target apps recommendation, called \modelnameneusa.
    \item It compares the performance of \modelnameneusa to state-of-the-art target apps recommendation.
    \item It describes the new dataset that we have collected and annotated for target apps recommendation, which we will make publicly available for research purposes.
    \item It includes more analysis of the collected data and the experimental results.
    \item It provides more details on our proposed context-aware target apps selection model \modelname.
\end{itemize}

This paper demonstrates that both our proposed models are able to outperform the state-of-the-art. Also, it provides new analysis and insights into the effect of context in both target apps selection and recommendation tasks. Finally, the joint analysis of context allows the reader to observe and compare the effectiveness of analyzing and incorporating user behavior data into the prediction.

The remainder of the paper is organized as follows. Section~\ref{sec:rel} provides a brief overview of the relevant studies in the literature. Section~\ref{sec:col} elaborates on our effort for collecting the data, followed by Section~\ref{sec:anl} where we analyze the collected data in depth. Then, in Sections~\ref{sec:method} and \ref{sec:nextapp} we describe both our proposed models for context-aware target apps selection and recommendation, respectively. Section~\ref{sec:expsetup} then includes details on the experimental setup, followed by Section~\ref{sec:res} discussing and analyzing the results. Finally, Section~\ref{sec:conclusion} concludes the paper and discusses possible future directions that stem from this work.

\section{Related Work}
\label{sec:rel}

Our work is related to the areas of mobile IR, context-aware search, target apps recommendation, human interaction with mobile devices (mobile HCI), and proactive IR. Moreover, relevant related research has been carried out in the areas of federated search and aggregated search and query classification.
In the following, we briefly summarize the related research in each of these areas.

A mobile IR system aims at enabling users to carry out all the classical IR operations on a mobile device \cite{DBLP:series/sbcs/CrestaniMS17}, as the conventional Web-based approaches fail to satisfy users' information needs on mobile devices~\cite{DBLP:conf/mhci/ChurchSBC08}. 
Many researchers have tried to characterize the main differences in user behavior on different devices throughout the years. 
In fact, \citet{DBLP:conf/www/SongMWW13} found significant differences in search patterns done using iPhone, iPad, and desktop.
Studying search queries is one of the main research topics in this area, as queries are one of the main elements of a search session.
\citet{DBLP:conf/chi/KamvarB06} conducted a large-scale mobile search query analysis, finding that mobile search topics were less diverse compared to desktop search queries.
Analogously, 
\citet{DBLP:conf/sigir/Guy16} and \citet{DBLP:journals/jasis/CrestaniD06} conducted comparative studies on mobile spoken and typed queries showing that spoken queries are longer and closer to natural language. All these studies show that significant changes in user behavior are obvious. Change of the interaction mode, as well as the purpose and change of the information need, are among the reasons for this change~\cite{AliannejadiSigir18}.

Moreover, there has been studies on how mobile search sessions compare with desktop search sessions~\cite{DBLP:journals/cacm/HalveyKS06,DBLP:conf/chi/HalveyKS06,DBLP:conf/sigir/BeitzelJCGF04,DBLP:conf/chi/BerkelLAFGHK16}.
\citet{DBLP:conf/chi/BerkelLAFGHK16} did a comprehensive analysis on how various inactivity gaps can be used to define an app usage session on mobile devices where they concluded that ``researchers should ignore brief gaps in interaction.'' 
\citet{DBLP:conf/chi/CarrascalC15} studied user interactions with respect to mobile apps and mobile search, finding that users' interactions with apps impact search. Also, they found that mobile search session and app usage session have significant differences.

Given that mobile devices provide rich contextual information about users' whereabouts, a large body of research has tried to study the effect of such information on users' behavior. 
\citet{DBLP:conf/mhci/ChurchO11} did a diary and interview study to understand users' mobile Web behavior.
\citet{DBLP:conf/chiir/AliannejadiHCPC19} conducted a field study where the recruited participants completed various search tasks in predefined time slots. They found that the temporal context, as well as the user's current activity mode (e.g., walking vs. sitting), influenced their perception of task difficulty and their overall search performance.

Also, \citet{DBLP:conf/chi/SohnLGH08} conducted a diary study in which they found that contextual features such as activity and time influence 72\% of mobile information needs. This is a very important finding, as it implies that using such information can greatly impact system performance and user satisfaction.
In fact, research on proactive IR mainly focuses on this fact~\cite{DBLP:conf/sigir/ShokouhiG15,DBLP:conf/wsdm/BenetkaBN17}.
\citet{DBLP:conf/sigir/ShokouhiG15} analyzed user interactions with information cards and found that the usage patterns of the proactive information cards depend on time, location, and the user's reactive search history. Proactive IR is very useful in a mobile context, where the user has a limited attention span for the mobile device and the applications running on it. Similarly, \citet{DBLP:conf/wsdm/BenetkaBN17} studied how various types of activities affect users' information needs. They showed that not only information needs vary across activities, but they also change during an activity. Our work follows a similar line leveraging the changing context to determine the target apps for a given query.

Other works focused on a more comprehensive comparison of user behavior where they found using information from user search sessions among different platforms can be used to improve performance~\cite{DBLP:conf/cikm/MontanezWH14}. It has also been shown that using external information such as online reviews can be used to improve the performance of search on mobile devices~\cite{Park:2015}.
\citet{DBLP:conf/cikm/ParkFLZ16} inferred users' implicit intentions from social media for the task of app recommendation.
This last work is closely related to our previous work \cite{AliannejadiSigir18} where we introduced the need for a unified mobile search framework as we collected cross-app queries through crowdsourcing. In contrast, we collect real-life cross-app queries over a longer period with an in situ study design in this work. 

Research on unified mobile search has considerable overlap with federated, aggregated search, and query classification. While federated search systems assume the environment to be uncooperative and data to be homogeneous, aggregated search systems blend heterogeneous content from cooperative resources~\cite{DBLP:journals/ftir/Arguello17}. Target apps selection, on the other hand, assumes an uncooperative environment with heterogeneous content.
Federated search has a long history in IR for Web search. In the case of uncooperative resources 
\citet{DBLP:journals/tois/CallanC01} proposed a query-based sampling approach to \emph{probe} the resources. 
\citet{Markov2014} carried out an extensive theoretical, qualitative, and quantitative analysis of different resource selection approaches for uncooperative resources. 
One could study probing for unified mobile search; however, we argue that apps could potentially communicate more cooperatively, depending on how the operating system would facilitate that. More recently, research on aggregated search has gained more attention. Aggregated search share certain similarities with target apps selection in dealing with heterogeneous data~\cite{DBLP:journals/ftir/ShokouhiS11}. However, research on aggregated search often enjoys fully cooperative resources as the resources are usually different components of the bigger search engine. For example, \citet{DBLP:conf/wsdm/Diaz09} proposed modeling the query dynamics to detect news queries for integrating the news \emph{vertical} in SERP.
Research on query classification has also been of interest for a long time in the field of IR.
Different strategies are used to assign a query to predefined categories. As mobile users are constantly being distracted by external sources, the queries often vary a lot, and it is not easy to determine if a query is related to the same information need that originated the previous query.
\citet{DBLP:conf/sigir/KangK03} defined three types of queries, each of which requiring the search engine to handle differently. \citet{DBLP:conf/sigir/ShenSYC06} introduced an intermediate taxonomy used to classify queries to specified target categories. \citet{DBLP:conf/sigir/CaoHSJSCY09} leveraged conditional random fields to incorporate users' neighboring queries in a session as context. More recently, \citet{Zamani:2016} studied word embedding vectors for the query classification task and proposed a formal model for query embedding estimation.

Predicting app usage has been studied for a long time in the field. Among the first works that tried to model app usage, \citet{DBLP:conf/icdm/LiaoLSLP12} proposed an app widget where users would see a list of recommended apps. Their model predicted the list of apps based on temporal usage profiles of users. Also, \citet{DBLP:conf/huc/HuangZMC12} studied different prediction models on this problem, including linear and Bayesian, where they found that contextual information, as well as sequential usage data, play important roles for accurate prediction of app usage. 
As smartphones kept evolving throughout these years, more data about various apps and users' context became available. As a result, more research focused on studying the effect of such information, as well as incorporating them into prediction models.
For instance, \citet{DBLP:conf/grc/LuLC14} studied the effect of location data and proposed a model that takes into account GPS data together with other contextual information.  
\citet{DBLP:conf/wsdm/Baeza-YatesJSH15} studied next app recommendation for improved home screen usage experience, extracting a set of personal and contextual features in a more commercial setting. \citet{DBLP:conf/huc/LeeLJJS16} found that the usage probabilities of apps follow the Zipf's law, as opposed to ``inter-running'' and running times which follow log-normal distributions. \citet{DBLP:conf/huc/WangYSZXLC16} modeled the apps following the idea of collaborative filtering, proposing a context-aware collaborative filtering model to unload and pre-load apps.
\citet{DBLP:conf/wasa/XuLZGZZZS18} modeled the sequential app usage using recurrent networks.
\citet{DBLP:conf/icde/ZhaoLJWXLY019} proposed the AppUsage2Vec model, inspired by doc2vec. Their proposed architecture includes an app-attention mechanism and a dual-DNN layer. 

As indicated in the literature, contextual and personal information have great impact in predicting user behavior on mobile devices. Also, researchers in the areas of federated and aggregated search have shown that contextual information play an important role in improved performance. In this work, we explore various sources of contextual information for both tasks. We also explore the use of recent app usage data as an implicit source of contextual information for target apps selection and show that it indeed provide useful contextual information to the model. Moreover, we study the collected data for both tasks, aiming to shed more light on the task of target apps selection and recommendation.

\section{Data Collection}
\label{sec:col}

In this section, we describe how we collected \textit{\istas} (\textbf{I}n \textbf{S}i\textbf{T}u collection of cross-\textbf{A}pp mobile \textbf{S}earch), which is, to the best of our knowledge, the first in situ dataset on cross-app mobile search queries. We collected the data in 2018 by recruiting 255 participants. The participants installed a simple Android app, called \app, for at least 24 hours on their smartphones. We asked them to use \app~to report their real-life cross-app queries as well as the corresponding target apps.
We first describe the characteristics of \app. Then, we provide details on how we recruited participants as well as the details on how we instructed them to report queries through the app. Finally, we give details on how we checked the quality of the collected data.

\begin{figure}
    \centering
    \includegraphics[width=\textwidth]{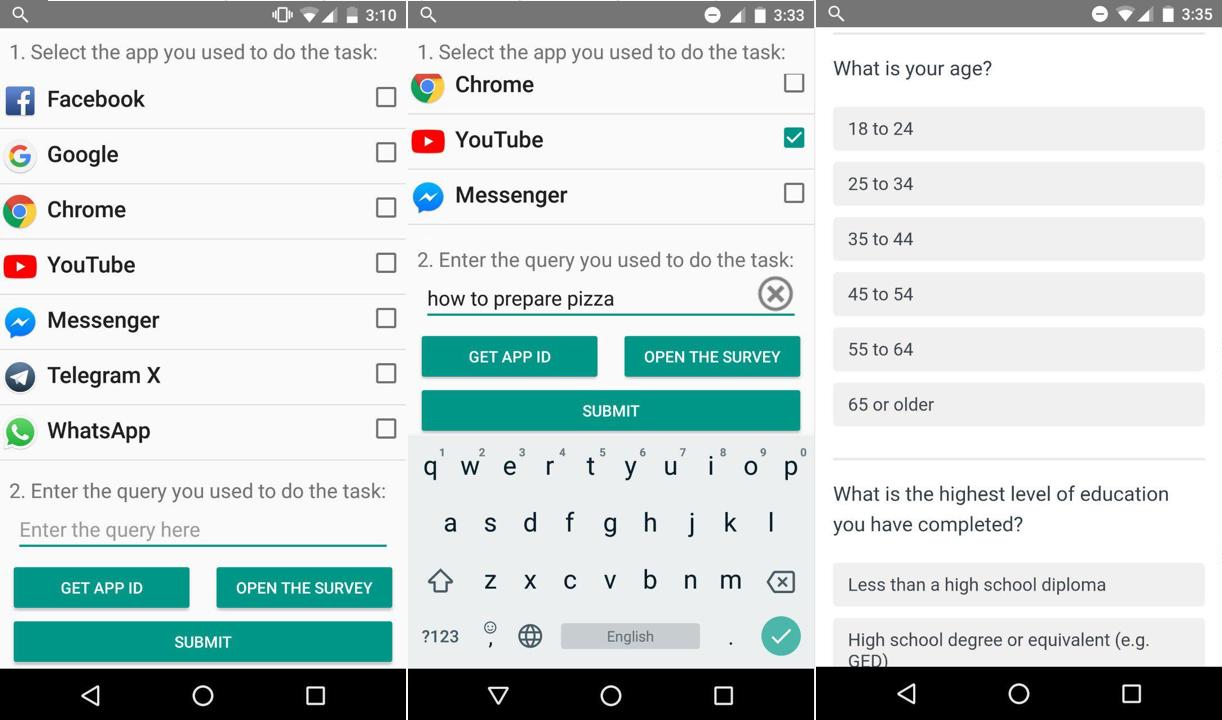}
    \caption{\app~interface on LG Google Nexus 5 as well as the survey. Checkboxes are used to indicate the target app for a query.}
    \label{fig:app-ui}
\end{figure}

\subsection{\app}
In order to facilitate the query report procedure, we developed \app, an Android app shown in \autoref{fig:app-ui}. We chose the Android platform because, in comparison with iOS, it imposes less restrictions in terms of sensor data collection and background app activity.

\partitle{User interface.} As shown in \autoref{fig:app-ui}, \app~consists of three sections. The upper part lists all the apps that are installed on the phone, with the most used apps ranked higher. The participants were supposed to select the app in which they had carried out their real-life search (e.g., \appname{Facebook}). In the second section, the participants were supposed to enter exactly the same query that they had entered in the target app (e.g., \appname{Facebook}). Finally, the lower part of the app, provided them easy access to a unique ID of their device and an online survey on their demographics and backgrounds. 

\partitle{Collected data.} Apart from the participants' input data, we also collected their interactions within \app~(i.e., taps and scrolling). Moreover, a background service collected the phone's sensors data. We collected data from the following sensors:
\begin{inlinelist}
    \item GPS;
    \item accelerometer;
    \item gyroscope;
    \item ambient light;
    \item WiFi; and
    \item cellular.
\end{inlinelist} 
Also, we collected other available phone data that can be used to better understand a user's context. The additional collected data are as follows:
\begin{inlinelist}
    \item battery level;
    \item screen on/off events;
    \item apps usage statistics; and
    \item apps usage events.
\end{inlinelist} 
Note that apps usage statistics indicate how often each app has been used in the past 24 hours, whereas apps usage events provides more detailed app events.\footnote{\url{https://developer.android.com/reference/android/app/usage/package-summary}} Apps usage events record user interactions in terms of:
\begin{inlinelist}
    \item launching a specific app;
    \item interacting with a launched app;
    \item closing a launched app;
    \item installing an app; and
    \item uninstalling an app;
\end{inlinelist}
The background service collected the data at a predefined time interval.
The data was securely transferred to a cloud service.

\subsection{Collection Procedure}
We recruited participants through an open call on Amazon Mechanical Turk.\footnote{\url{http://mturk.com}} 
\revision{The study received the approval of the ethics committee of the university. We provided a clear statement to the participants about the kind of data that we were collecting and the purpose of the study. Furthermore, we used secure encrypted servers to store users' data.}
We asked the participants to complete a survey inside \app. Moreover, we mentioned all the steps required to be done by the participants in order to report a query. 
In short, we asked the participants to open \app~after every search they did using any installed app on their phones. Then, we asked them to report the app as well as the query they used to perform their search task. We encouraged the participants to report their search as soon as it occurs, as it was very crucial to capture their context at the right moment. 

After running several pilot studies, over a period of 12 weeks we recruited 255 participants, asking them to let the app running on their smartphones for at least 24 hours and report at least 5 queries. Since some people may not submit 5 search queries during the period of 24 hours, we asked them to keep the app running on their phones after the first 24 hours until they report 5 queries. Also, we encouraged them to continue reporting more than 5 queries for an additional reward. As incentive, we paid the participants \$0.2 per query. We recruited participants only from English-speaking countries.

\subsection{Quality Check}
During the course of data collection, we performed daily quality checks on the collected data. The checks were done manually with the help of some data visualization tools that we developed. 
\revision{We visualized the use of selected apps in the participant's app-usage history in a timeline to validate a user's claim when they report using a specific app for their search.}
As we were paying participants a reward per query, we carefully studied the submitted queries as well as user interactions to prevent participants from reporting false queries. For each query, we checked the apps usage statistics and events for the same day. If a participant reported a query in a specific app (e.g., \appname{Facebook}) but we could not find any recent usage events regarding that app, we assumed that the query was falsely reported.
Moreover, if a participant reported more than 10 queries per day, we took some extra quality measures into account.
Finally, we approved 6,877 queries out of 7,750 reported queries.

\subsection{Data Transfer}
To prevent unwanted career charges, we limited the data transfer to WiFi only. For this reason, we provided a very flexible implementation to manage the data in our app. In our app design, the data is stored locally as long as the device is not connected to a WiFi network. As soon as a WiFi connection is available, the app uploads the data to the cloud server. We made this point very clear in the instructions and asked the participants to take part in the study only if they had a strong WiFi connection at home or office.

\subsection{Privacy Concerns}
Before asking for required app permissions, we made clear statement about our intentions on how we were going to use the participants' collected data as well as what was collected from their devices. We ensured them that their data were stored on secure cloud servers and that they could opt out of the study  at any time. In that case we would remove all their data from the servers. While granting apps usage access was mandatory, granting location access was optional. We asked participants to allow \app~access their locations only if they felt comfortable with that. Note that, through the background service, we did not collect any other data that could be used to identify participants.

\section{Data Analysis}
\label{sec:anl}
In this section, we describe the basic characteristics of \istas, and present a thorough analysis of target apps, queries, sessions, and context.

\begin{figure}
    \centering
    \includegraphics[width=0.75\textwidth]{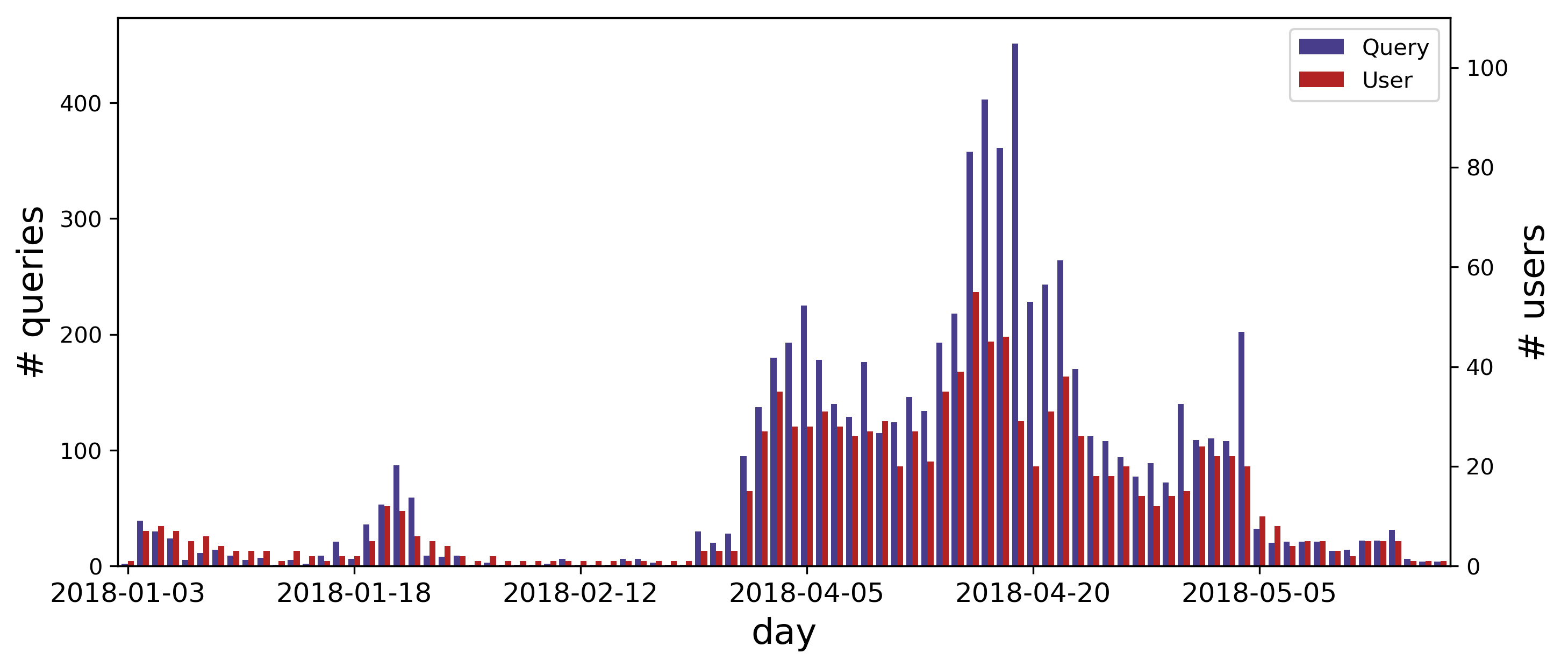}
    \caption{Number of queries and active participants per day, during the course of data collection (best viewed in color).}
    \label{fig:daily-queries}
\end{figure}

\subsection{Basic Statistics}
\partitle{\istas.}
During the period of 86 days, with the help of 255 participants, we collected 6,877 search queries and their target apps as well as sensor and usage data. The collected raw data was over 300~gigabytes.
Here, we summarize the main characteristics of the participants based on the submitted surveys. Over 59\% of the participants were female. Nearly 50\% of them were aged between 25-34, followed by 22\% between 35-44, and 15\% 18-24 years. Participants were from all kinds of educational backgrounds ranging from high school diploma to PhD. In particular, 32\% of them had a college degree, followed by 30\% with a bachelor's degree. Smartphone was the main device used for connecting to the Internet for 53\% of the participants, followed by laptop (25\%). Among the participants, 67\% used their smartphones more often for personal reasons rather than for work. Finally, half of the participants stated that they use their smartphones 4 hours a day or more.
Table~\ref{tab:stats} lists basic characteristics of \istas. In the table we see that we received $26.97 \pm 50.21$ queries per user. Notice that the high value of standard deviation ($=50.21$) is due to existence of some users who submitted numerous queries.
Moreover, \autoref{fig:daily-queries} shows the number of queries and active participants per day during the data collection period. Note that, as shown in \autoref{fig:daily-queries}, in the first half collection period, we were mostly developing the visualization tools and did not recruit many participants.

\partitle{\lsapp.}
We collected \lsapp (\textbf{L}arge dataset of \textbf{S}equential mobile \textbf{App} usage) using the \app\footnote{\url{https://github.com/aliannejadi/uSearch}} data collection tool during an eight-month period involving 292 users. 
Notice that 255 of the users were the same people that were involved in collecting \istas. The extra 37 participants were the ones that either did not submit any queries during this period, or submitted low-quality queries and were removed in the quality check phase.
Table~\ref{tab:stats:lsapp} summarizes the statistics of \lsapp. Since we observed many repeated app usage records with very short differences in time (< 10 seconds), we considered all repeated app usage records with less than one minute time difference as one record. 
\revision{Also, as the app usage data includes various system apps, we filtered out background system packages and kept only the most popular apps in the data. We identify the most popular apps based on the data we collected in this dataset.}

\begin{table}[t]
    \centering
    \caption{Statistics of \istas.}
    \label{tab:stats}
    \begin{tabular}{ll}
        \toprule
         \# queries & 6,877 \\
         \# unique queries & 6,262 \\
         \# users & 255 \\
         \# unique apps & 192 \\
         \# search sessions & 3,796 \\
         \# days data collected & 86 \\
         Mean queries per user & 26.97 $\pm$ 50.21 \\
         Mean queries per session & 1.81 $\pm$ 2.88 \\
         Mean queries per day & 79.96 $\pm$ 101.27 \\
         Mean days of report per user & 7.38 $\pm$ 15.95 \\
         Mean unique apps per user & 5.14 $\pm$ 14.06 \\
         Mean query terms & 3.00 $\pm$ 1.96 \\
         Mean query characters & 17.53 $\pm$ 10.46 \\
         \bottomrule
    \end{tabular}
\end{table}

\begin{table}[t]
    \centering
    \caption{Statistics of \lsapp.}
    \label{tab:stats:lsapp}
    \begin{tabular}{ll}
        \toprule
         \# app usage records & 599,635\\
         \# sessions & 76,247\\
         \# unique apps & 87 \\
         \# users &  292\\
         Mean duration/user & 15 days\\
         \midrule
         \revision{Mean session time length} & 5:26 \\
         \revision{Median session time length} & 1:46 \\
         \revision{Mean unique apps in each session} & 2.18 \\
         \revision{Median unique apps in each session} & 2 \\
         \revision{Mean app switches within a session} & 5.46 \\
         \revision{Median app switches within a session} & 2 \\
         \midrule
         \# train instances &  464,903\\
         \# validation instances & 66,300 \\
         \# test instances &  132,751\\
         \bottomrule
    \end{tabular}
\end{table}

\subsection{Apps}

\begin{figure}
    \centering
    \includegraphics[width=0.75\textwidth]{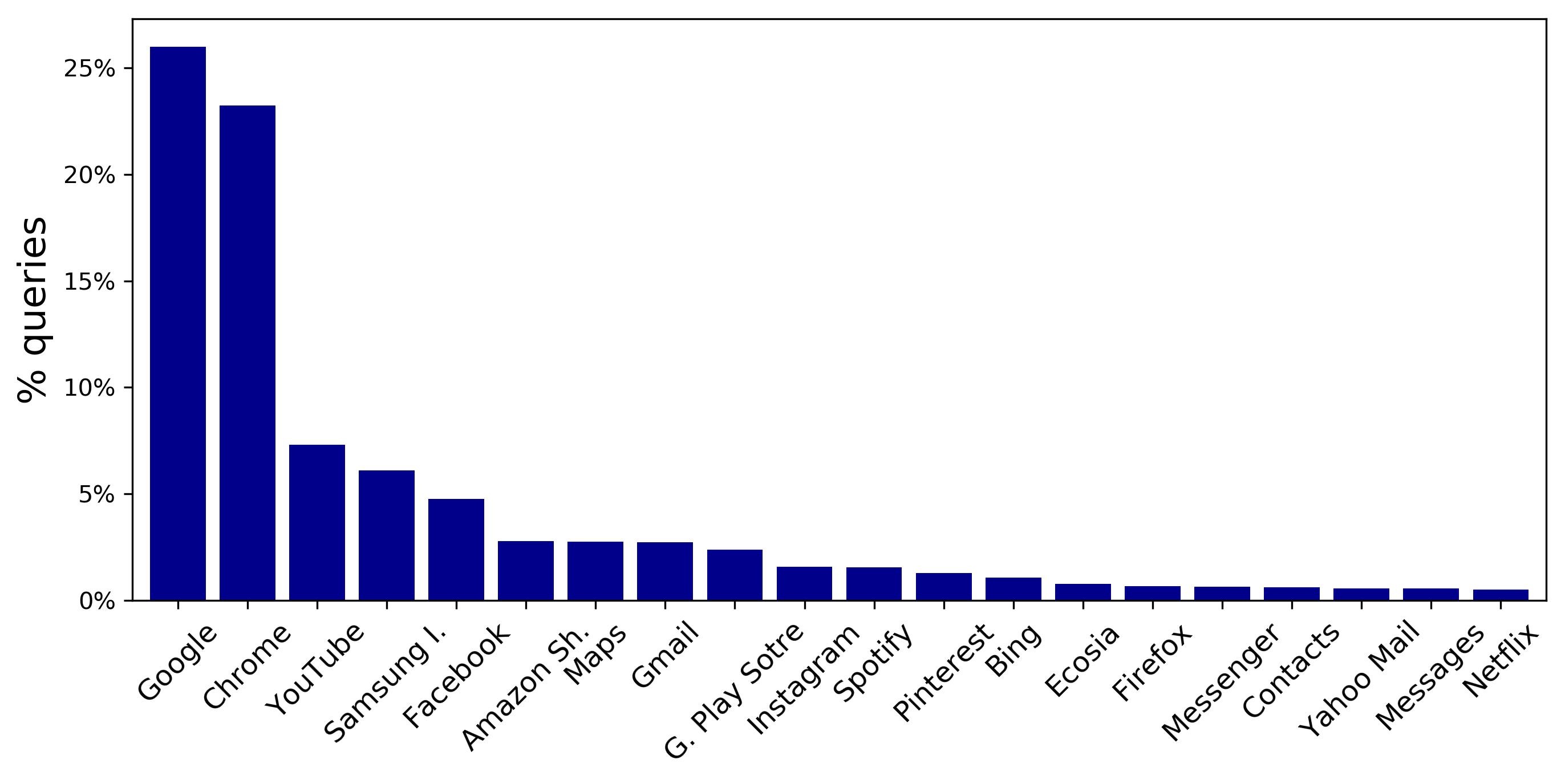}
    \caption{Number of queries per app for top 20 apps in \istas.}
    \label{fig:app_hist}
\end{figure}

\partitle{How apps are distributed.} 
Figure~\ref{fig:app_hist} shows how queries are distributed with respect to the top 20 apps. 
We see that the top 20 apps account for 88\% of the searches in \istas, showing that the app distribution follows a power-law.
While \appname{Google} and \appname{Chrome} queries respectively attract 26\% and 23\% of the target apps, users conduct half (51\%) of their search tasks using other apps. This finding is inline with what was shown in a previous work~\cite{AliannejadiSigir18}, even though we observe a higher percentage of searches done using \appname{Google} and \appname{Chrome} apps. 
In~\cite{AliannejadiSigir18}, we collected a dataset cross-app queries called UniMobile under a different experimental setup where we asked the participants to submit cross-app queries for given search tasks. 
Therefore, the differences in the collected data can be due to two reasons:
\begin{inlinelist}
    \item \istas~is collected in situ and on mobile devices, thus being more realistic than UniMobile;
    \item \istas~queries reflect real-life information needs rather than a set of given search tasks, hence the information need topics are more diverse than UniMobile.
\end{inlinelist}
Moreover, we observe a notable variety of apps among the top 20 apps, such as \appname{Spotify} and \appname{Contacts}.
We also see \appname{Google Play Store} among the top target apps. This suggests that people use their smartphones to search for a wide variety of information, most of which were done with apps other than \appname{Google} or \appname{Chrome}. It should also be noted that users seek the majority of their information needs on various apps, even though there exists no unified mobile search system on their smartphones, suggesting that they might even do a smaller portion of their searches using \appname{Google} or \appname{Chrome}, if a unified mobile search system was available on their smartphones. 

\begin{figure}
\vspace{-0.4cm}
    \centering
    \subfloat[apps per user]{
        \includegraphics[width=0.4\textwidth]{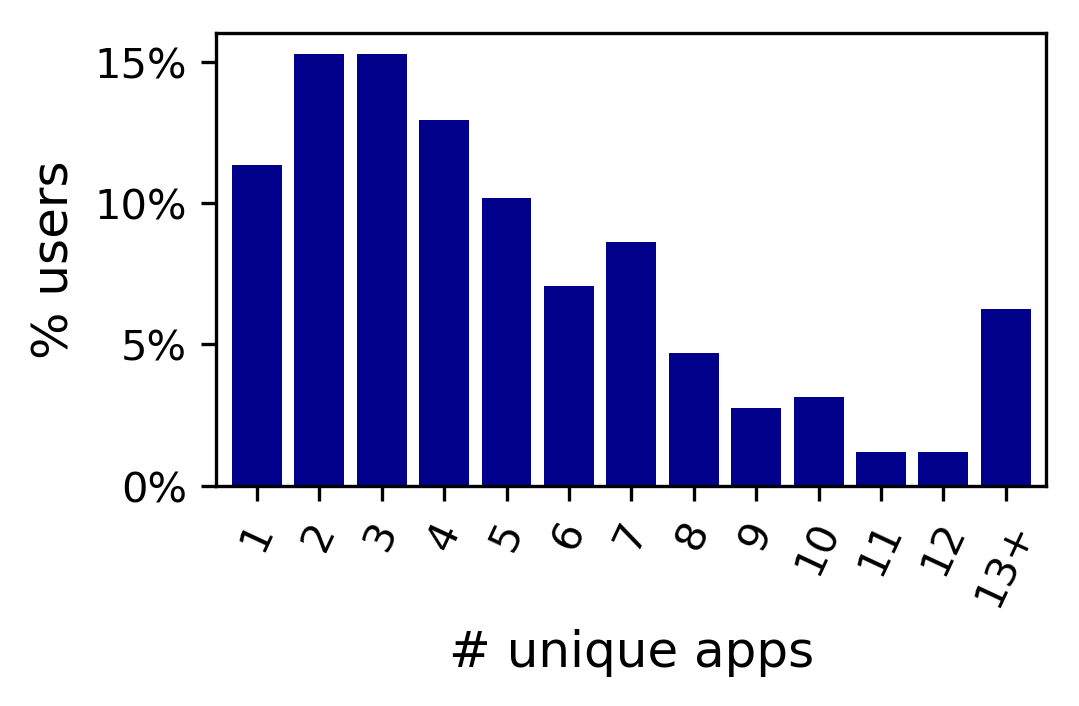}
        \label{fig:user_unique_app}
    }
    \subfloat[apps per session]{
        \includegraphics[width=0.4\textwidth]{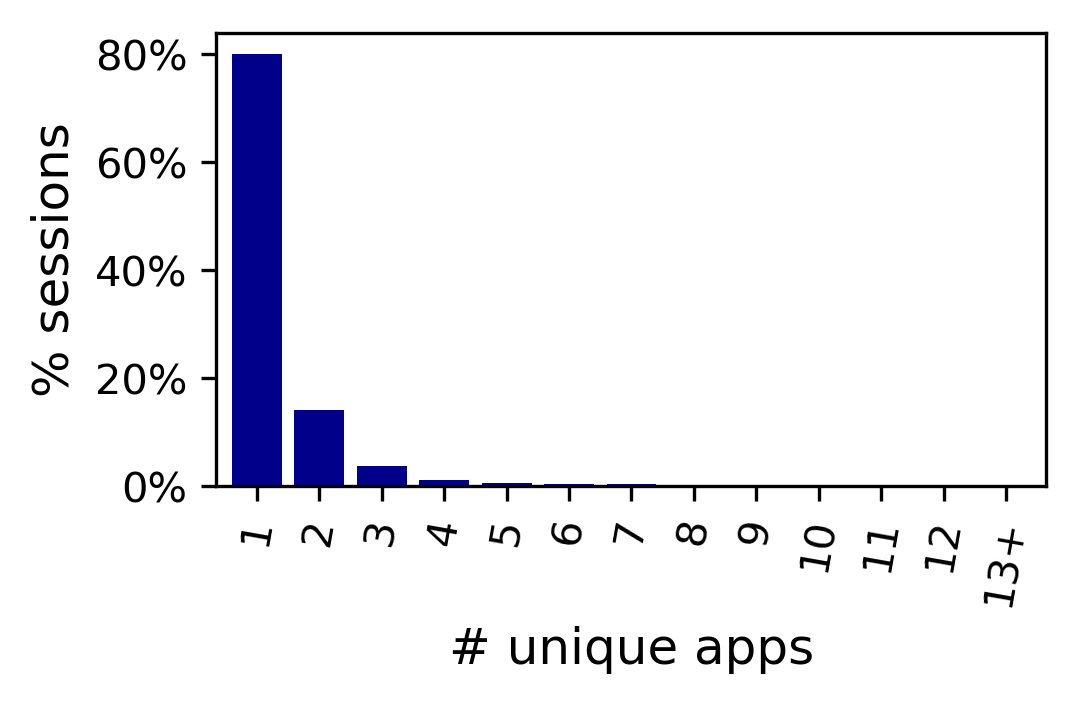}
        \label{fig:session_unique_app}
    }
    \caption{Distribution of unique apps per user and per session in \istas.}
    \label{fig:unique_apps}
\end{figure}

\partitle{How apps are selected.} Here, we analyze the behavior of the participants in \istas, as they searched for real-life information needs, in terms of the apps they chose for performing the search. Figure \ref{fig:user_unique_app} shows the distribution of unique apps per user. We can see how many users selected a certain number of unique apps, with an average of 5.14 unique apps per user. Again, this indicates that users seek information in a set of diverse apps. It is worth noting that in Figure~\ref{fig:user_unique_app}, we observe a totally different distribution compared to \cite{AliannejadiSigir18}, where the average number of unique apps per user was much lower. We believe this difference is due to the fact that the participants in our work reported their real-life queries, as opposed to the crowdsourcing setup of \cite{AliannejadiSigir18}.

On the other hand, Figure~\ref{fig:session_unique_app} plots the distribution of unique apps with respect to sessions, which is how many unique apps were selected during a single search session. We see an entirely different distribution where the average number of unique apps per task is $1.36$. This shows that while users seek information using multiple apps, they are less open to switching between apps in a single session. This can partly be due to the fact that switching between apps is not very convenient. However, this behavior requires more investigation to be fully understood, that we leave for future work. 

\begin{table}[]
    \centering
    \caption{Cross-app query attributes for 9 apps. The upper part of the table lists the distribution of number of query terms as well as mean query terms per app. The lower part lists the query overlap at different similarity thresholds (denoted by $\tau$) per app. \textit{All} shows query distributions across all apps.}
    \begin{tabular}{lccccccccccccccccccccccccccc}
        \toprule
        & \rotatebox{90}{All} & \rotatebox{90}{Google} & \rotatebox{90}{YouTube} & \rotatebox{90}{Facebook} & \rotatebox{90}{Amazon Sh.} & \rotatebox{90}{Maps} & \rotatebox{90}{Gmail} & \rotatebox{90}{G. Play Store} & \rotatebox{90}{Spotify} & \rotatebox{90}{Contacts} \\
        \midrule \midrule
         \multicolumn{2}{l}{\# terms} & \multicolumn{8}{c}{\textbf{Query term distribution}} \\
         \midrule
        1 & 22\% & 13\% & 11\% & 22\% & 12\% & 25\% & 57\% & 49\% & 29\% & 81\% \\
        2 & 28\% & 26\% & 29\% & 48\% & 45\% & 27\% & 30\% & 33\% & 35\% & 10\% \\
        3 & 20\% & 21\% & 24\% & 16\% & 25\% & 18\% & 9\% & 12\% & 24\% & 7\% \\
        4 & 12\% & 13\% & 18\% & 10\% & 10\% & 13\% & 3\% & 4\% & 7\% & 2\% \\
        > 4 & 17\% & 26\% & 17\% & 4\% & 10\% & 17\% & 1\% & 1\% & 6\% & 0\% \\
        \midrule
        Mean & 3.00 & 3.49 & 3.19 & 2.34 & 2.74 & 3.07 & 1.61 & 1.75 & 2.31 & 1.31 \\
        \midrule \midrule
        \multicolumn{1}{c}{$\tau$} && \multicolumn{8}{c}{\textbf{Query overlap}} \\
        \midrule
        > 0.25 & 56\% & 39\% & 41\% & 28\% & 27\% & 26\% & 27\% & 25\% & 8\% & 14\% \\ 
        > 0.50 & 19\% & 11\% & 15\% & 13\% & 7\% & 11\% & 12\% & 12\% & 4\% & 10\% \\
        > 0.75 & 13\% & 5\% & 8\% & 11\% & 5\% & 9\% & 12\% & 10\% & 2\% & 10\% \\
        \bottomrule
    \end{tabular}
    \label{tab:query-atts}
\end{table}

\subsection{Queries}
\label{sec:queries}
In order to understand the differences in user behavior while formulating their information needs using different apps, we conducted an analysis on the attributes of the queries with respect to their target apps. First, we start by studying the number of query terms in each app for the top 9 apps in \istas.

\partitle{How query length differs among apps.} 
The upper part of \autoref{tab:query-atts} lists the distribution of the number of query terms in the whole dataset (denoted by \textit{All}) as well as in each app. It also lists the average query terms per app. As we can see, the average query length is 3.00, which is slightly lower than previous studies on mobile query analysis~\cite{DBLP:conf/sigir/Guy16,DBLP:conf/chi/KamvarB06}. 
However, the average query length for apps that deal with general web search such as \appname{Google} is higher ($=3.49$). This indicates that users submit shorter queries to other apps. For instance, we see that \appname{Contacts} has the lowest average query length ($=1.31$), as its queries are mainly contact names. Also, \appname{Gmail} and \appname{Google Play Store} have an average query length lower than 2 as most searches are keyword based (e.g., part of an email subject or an app name) . This difference shows a clear behavioral difference in formulating queries using different apps. Moreover, we can see that the distribution of the number of query terms varies among different apps; take \appname{Contacts} as an example, whose single-term queries constitute 81\% of its query distributions, which are often names of user's personal contacts. 
This indicates that the structure of queries vary  across the target apps. Studying the most frequent query unigrams of each app also confirms this finding. For example, \appname{Google}'s most popular unigrams are mostly stopwords (i.e., ``to'', ``the'', ``of'', ``how''), whereas \appname{Facebook}'s most popular unigrams are not (i.e., ``art'', ``eye'', ``wicked'', ``candy'').

\partitle{How query similarity differs across apps.} 
The lower part of \autoref{tab:query-atts} lists the query similarity or query overlap using a simple function used in previous studies~\cite{DBLP:journals/tweb/ChurchSCB07,AliannejadiSigir18}. We measure the query overlap at various degrees and use the similarity function $\text{sim}(q_1, q_2) = |q_1 \cap q_2|/|q_1 \cup q_2|$, simply measuring the overlap of query terms. We see that among all queries, 18\% of them are similar to no other queries. We see a different level of query overlap in queries belonging to different apps. The highest overlap is among queries from Web search apps such as \appname{Chrome} and \appname{Google}. Lower query similarity is observed for personal apps such as \appname{Facebook} and for more focused apps such as \appname{Amazon Shopping}. Note that the query overlap is higher when all app queries are taken into account (All), as opposed to individual apps. 
This shows that users tend to use the same query or a very similar query when they switch between different apps,
suggesting that switching between apps is part of the information seeking or query reformulation procedure on mobile devices.

\subsection{Sessions}
\label{sec:session}

\partitle{\istas.}
A session is a ``series of queries by a single user made within a small range of time''~\cite{DBLP:journals/sigir/SilversteinHMM99}. Similar to previous work~\cite{DBLP:journals/sigir/SilversteinHMM99,DBLP:conf/chi/KamvarB06,DBLP:conf/chi/CarrascalC15}, we consider a five-minute range of inactivity as closing a session. \istas~consists of 3,796 sessions, with 1.81 average queries per session. The majority of sessions have only one query ($=66\%$). Similarly, as shown in Figure~\ref{fig:session_unique_app}, participants use only one app in the majority of sessions ($=80\%$). We also studied how similar queries were distributed among single-app sessions as compared to multiple-app sessions. We found that queries are similar to each other in multiple-app sessions. More specifically, query overlap at the threshold of > 0.25 is 49\% and 56\% in single-app and multiple-app sessions, respectively. This suggests that users tend to switch between apps to search for the same information need as they reformulate their queries.

\begin{figure}
    \centering
    \includegraphics[width=1.3\textwidth]{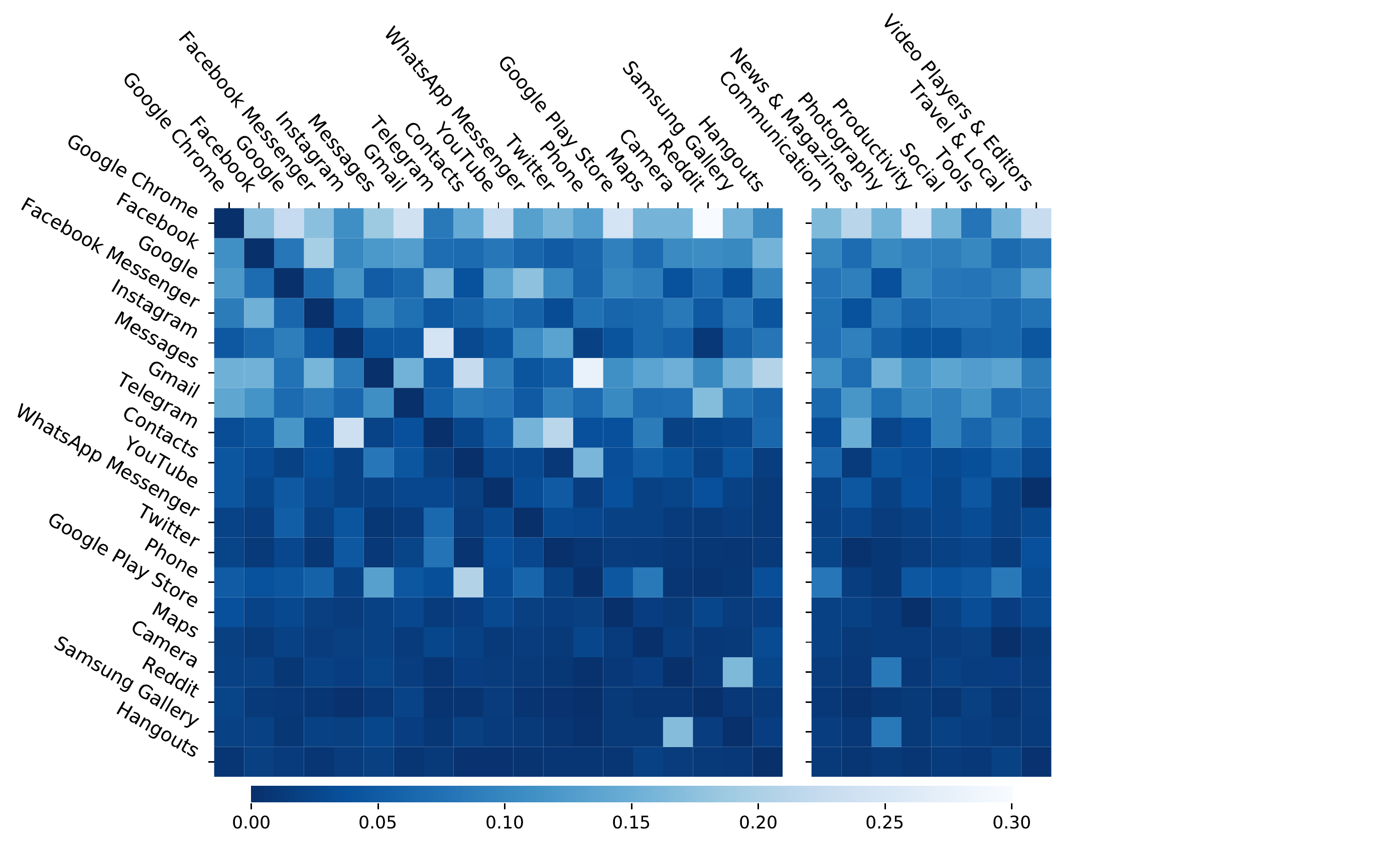}
    \caption{Heat map depicting co-occurrence of apps in same sessions with other apps in \lsapp. The graph on the left shows the co-occurrence at app level, whereas the one on the right shows it at category level. Popular apps such as \appname{Google Chrome} dominantly co-occur with most of other apps in various categories.}
    \label{fig:heatmap}
\end{figure}

\begin{figure}
    \subfloat[\revision{Switch pattern among app categories.}]{\includegraphics[width=1\textwidth]{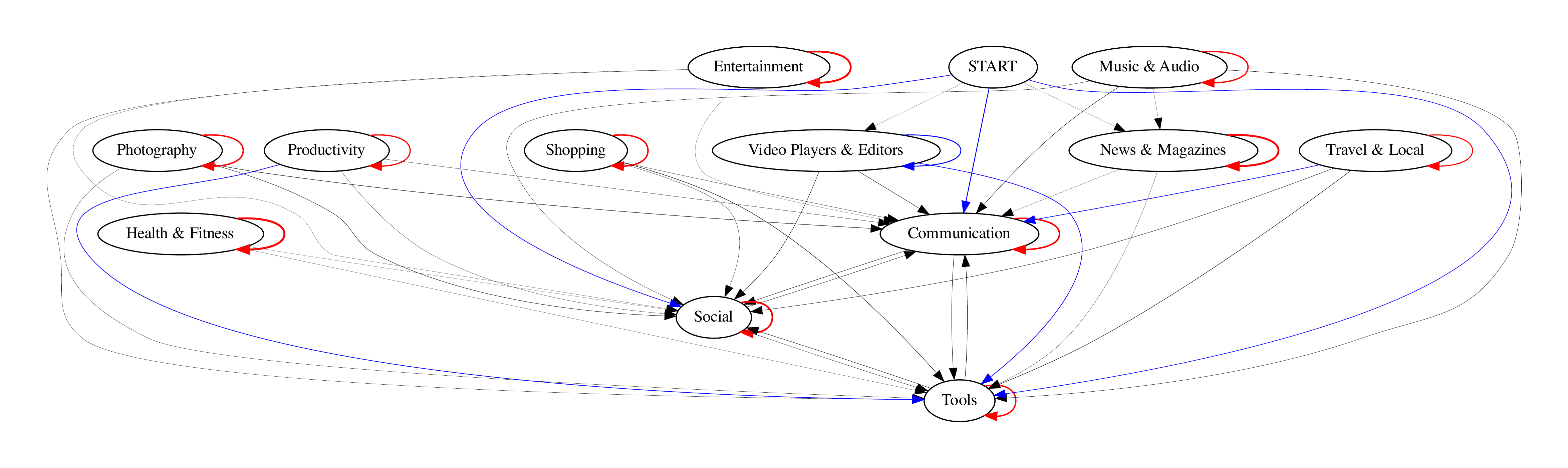} \label{fig:markov_chain_cat}}
    
    \subfloat[\revision{Switch pattern among apps belonging to Social and Communication categories.}]{\includegraphics[width=0.6\textwidth]{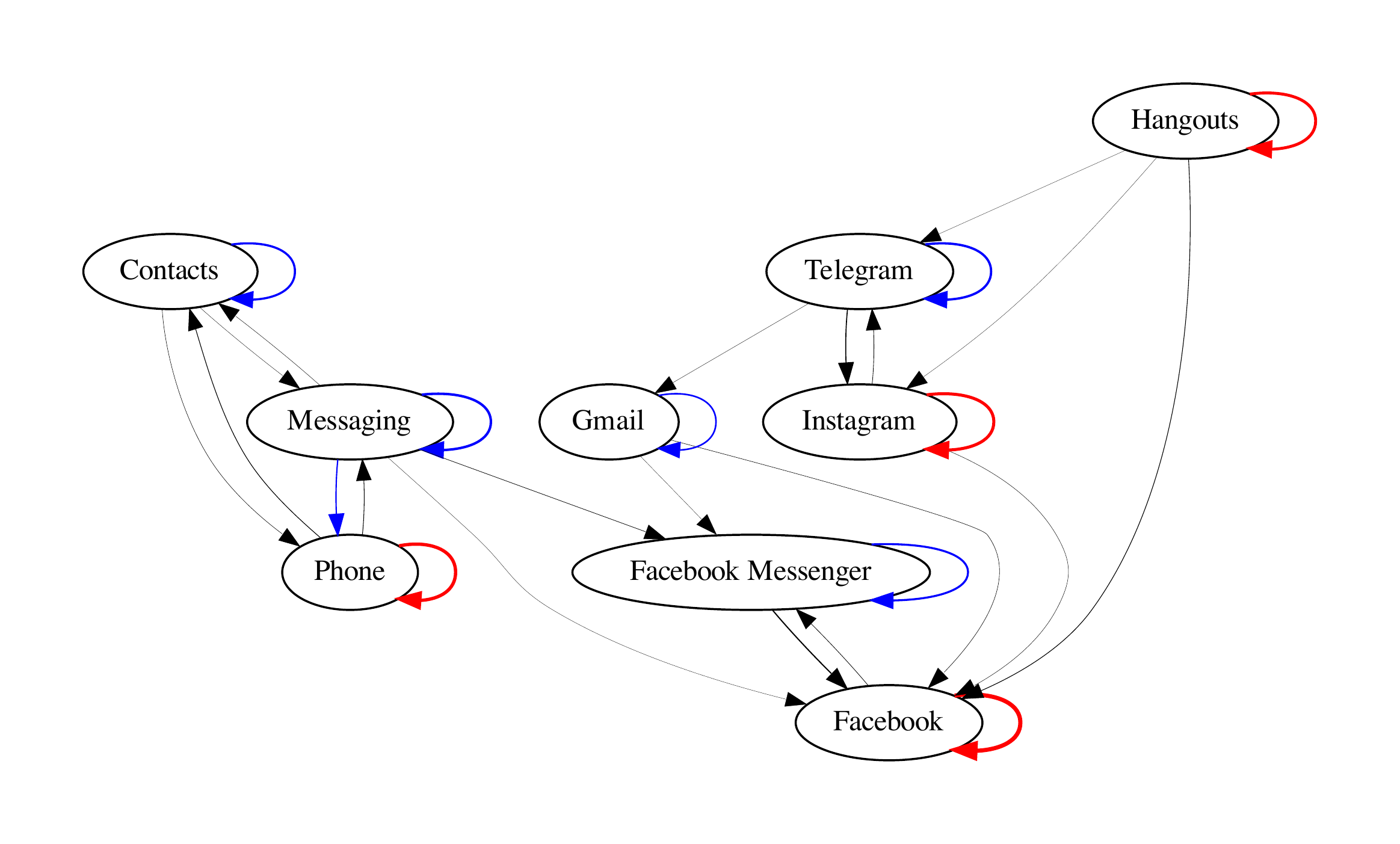} \label{fig:markov_chain_app}}
    \caption{\revision{App switch pattern in sessions with a Markov model on (a) app category level and (b) apps belonging to two categories. Edges represent a transition probability of over 0.05. Edges are directed and weighted by transition probability, with blue and red edges indicating over 0.2 and 0.4 transition probabilities, respectively.}}
    \label{fig:markov_chain}
\end{figure}

\partitle{\lsapp.}
For consistency with the search sessions, we consider a five-minute range of inactivity also for \lsapp. It is worth noting that even though the relevant work suggests smaller inactivity periods~\cite{DBLP:conf/chi/BerkelLAFGHK16,DBLP:conf/chi/CarrascalC15}, we assume that a session ends after five minutes of inactivity to tackle the noisy app usage data and appearance of background services while the user is continuously using the same app.
The collection contains a total number of 61,632 app usage sessions. 
\revision{Table \ref{tab:stats:lsapp} reports the mean and median length of sessions in terms of time, number of switches between apps. Also, we report the mean and median number of unique apps that users launch in a session. Comparing the number of app switches with unique apps, we see that in many sessions, users tend to work with two apps and do multiple switches between them. To gain more insight into the nature of app switches, we perform the two analyses shown in Figures \ref{fig:heatmap} and \ref{fig:markov_chain}.}

Our first goal here is to show how top-used apps in \lsapp are used in the same session by users. To this end, we count the number of co-occurrences in sessions and normalize the numbers by summing over all co-occurrence values. Note that we describe the definition of an app usage session in Section~\ref{sec:session}.
Figure~\ref{fig:heatmap} illustrates the co-occurrence values in the form of a heat map with other apps displayed based on individual apps on the left, as well as categories on the right. 
\revision{We have used the official taxonomy of apps from Google Play Store.}
Since every app always co-occurs with itself (hence having the maximum value of each row), we have set the diagonal values to zero for a better quality of the figure.
We see from the first column that \appname{Google Chrome} has the highest share of usage compared to other apps because it has the highest value of most rows. It is interesting to see that users employ more popular apps such as \appname{Google} together with the other apps in most of the sessions. As argued in \cite{AliannejadiSigir18}, users tend to use multiple apps to complete a single search task. Switching between popular search apps in our data suggests the same behavioral pattern is observed here. 
On the right side of the figure, we see how each app co-occurs with other apps based on their categories. It is interesting to observe that some app features could affect what type of apps co-occur. For example, observing the co-occurrences of the ``Photography'' app category, we see that social networking apps such as \appname{Instagram} and \appname{Telegram} exhibit some of the lowest co-occurrence values. This could be because of the photography features that already exist in such apps. Conversely, we see that apps such as \appname{Messages} and \appname{Gmail} co-occur more frequently. 
Also, we see that other apps belonging to the same or related categories are, in some cases, used in a session. For example, we see that \appname{Phone} co-occurs with \appname{Messaging} and \appname{Contacts}. It is also interesting to observe the lowest row of the figure, showing the co-occurrence of \appname{Hangouts}. We see that while \appname{Hangouts} exhibits high co-occurrence with social media apps like \appname{Facebook} and \appname{Instagram}, it is not highly used in the same sessions with instant messaging apps such as \appname{WhatsApp Messenger}, \appname{Facebook Messenger}, and \appname{Messages}. This suggests that apps that fall into the same high-level category (i.e., social networking) tend to co-occur in a session, as users achieve different goals. However, users tend to use only one of the apps that fulfill very similar needs (i.e., instant messaging).

\revision{We illustrate the transition probabilities between app categories in Figure~\ref{fig:markov_chain_cat}. The figure shows a Markov model of how users switch between apps that belong to different categories in a session. We see that the majority of sessions start with apps of Tools, Social, and Communication categories. Although users switch between various categories of apps, we see that they mostly tend to use apps of the same categories in a single session. This suggests that perhaps the types of tasks they complete in a single session can be done using a single or a set of apps with similar proposes (i.e., belong to the same category). To explore the transition probabilities between apps, we show in Figure~\ref{fig:markov_chain_app} a Markov model of app transitions in sessions for Social and Communication apps. Here, we also see that even though users switch often among different apps, there is a higher tendency to switch to the same app (i.e., blue- and red-colored edges indicate higher probabilities). This suggests that while users are trying to perform a task, they might be interrupted by environmental distractions or notifications on their phones, closing the current app and opening it later. In particular we see a self-transition probability of over 0.4 on \appname{Phone}, \appname{Instagram}, \appname{Hangouts}, and \appname{Facebook}. This is perhaps related to the users' tendency to engage with these apps for longer, leading to a higher probability of interruption. Interestingly, we observe that native Communication apps (i.e., \appname{Contacts}, and \appname{Phone}, \appname{Messaging}) form a cluster on the left side of the figure, with users switching mainly among the three apps while switching to other apps only through \appname{Messaging}.
}

\subsection{Context}
\label{sec:context}
\begin{figure}
\vspace{-0.3cm}
    \centering
    \includegraphics[width=0.75\textwidth]{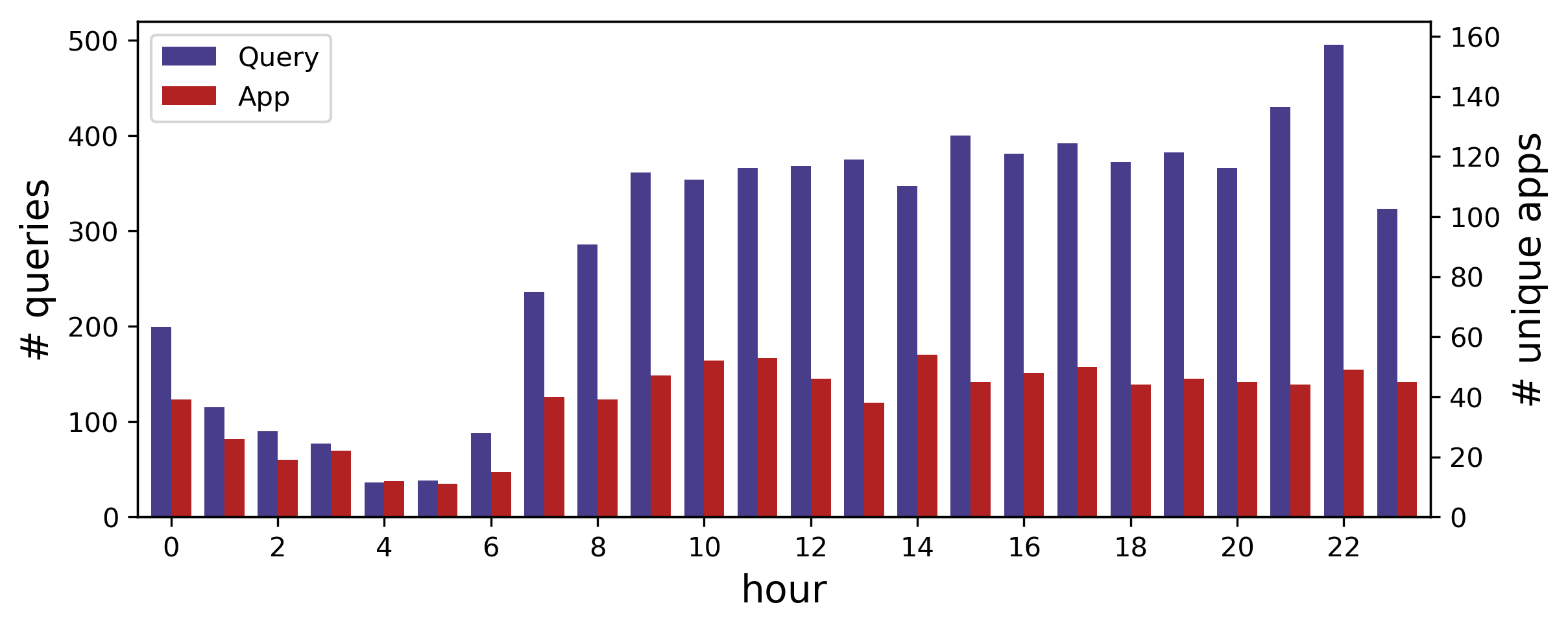}
    \caption{Time-of-the-day distribution of queries and unique apps (best viewed in color).}
    \label{fig:query_app_hour}
\end{figure}

\partitle{Temporal behavior.} We analyze the behavior of users as they search with respect to day-of-week and time-of-day.
We see that the distribution of queries on different days of week slightly peaks on Fridays. Notice that in this analysis, we only include the users that participated in our study for more than six days.
Moreover, Figure \ref{fig:query_app_hour} shows the distribution of queries and unique target apps across time-of-day for all participants. Our findings agree with similar studies in the field~\cite{DBLP:conf/sigir/BeitzelJCGF04,DBLP:conf/chi/HalveyKS06}. As we can see, more queries are submitted in the evenings, however we do not see a notable difference in the number of unique target apps.

\begin{figure}
    \includegraphics[width=0.75\textwidth]{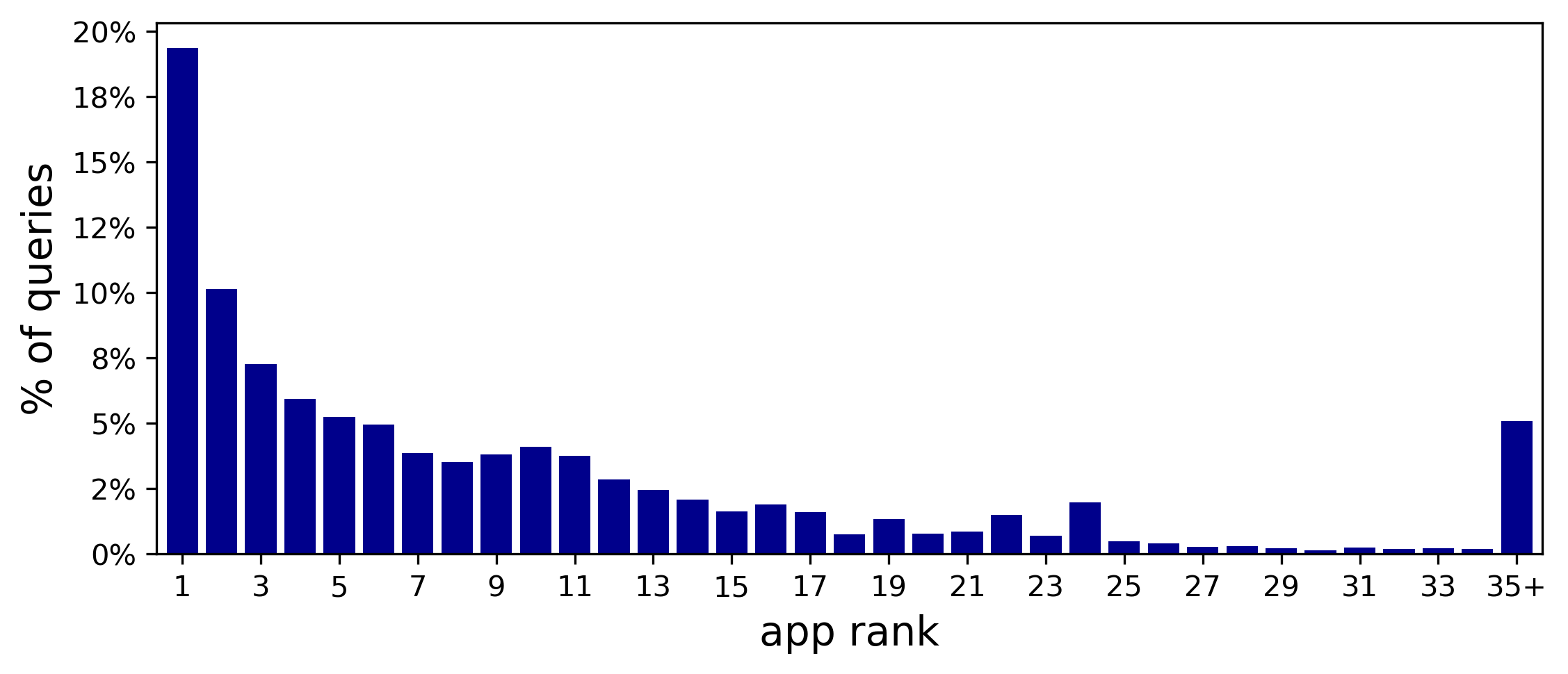}
        \caption{Apps usage context ranking distribution of relevant target apps. Lower values of x axis mean that the app has been used more often in the past 24 hours.}
    \label{fig:app_rank_hist}
\end{figure}

\partitle{Apps usage context.} 
We define a user's \textit{apps usage context} at a given time $t$ as the apps usage statistics of that specific user during the 24 hours before $t$. Apps usage statistics contain details about the amount of time users spent on every app installed on their smartphones. This gives valuable information on users' personal app preferences as well as their contexts. For example, a user who has interacted with travel guide apps in the past 24 hours is probably planning a trip in the near future. Therefore, we analyze how users' apps usage context can potentially help a target app selection model. Figure \ref{fig:app_rank_hist} shows the histogram of target app rankings in the users' apps usage contexts. 
We see that participants often looked for information in the apps that they use more frequently. For instance, 19\% of searches were done on the most used app, followed by 10\% on the second most used app.
We also see that, in most cases, as the ranking increases, the percentage of target apps decreases, suggesting that incorporating users app usage context is critical for target apps selection.

\section{Context-Aware Neural Target Apps Selection}
\label{sec:method}
In this section, we propose a context-aware neural model called \modelname (\textbf{C}ontext-aware \textbf{N}eural \textbf{T}arget \textbf{A}pps \textbf{S}election), which is an extension to our recent neural target apps selection model (i.e., NTAS1) \cite{AliannejadiSigir18}. Our model takes as input a query $q$, a candidate app $a$, and the corresponding query context $c_q$ and produces a score indicating the likelihood of the candidate app $a$ being selected by the user as the target app for the query $q$. In the following, we first describe a \emph{general} framework for context-aware target apps selection and further explain how it is implemented and how context is incorporated into the framework.

Formally, the \modelname framework estimates the probability $p(S=1 | q, a, c_q; A)$, where $S$ is a binary random variable indicating whether the app $a$ should be selected ($S=1$) or not ($S=0$). $A$ denotes the set of candidate apps. This set can be all possible apps, otherwise those that are installed on the user's mobile device, or again a set of candidate apps that is obtained by another model in a cascade setting. The app selection probability in the \modelname framework is estimated as follows:
\begin{equation}
    p(S=1 | q, a, c_q; A) = \psi(\phi_Q(q), \phi_A(a), \phi_C(c_q))~,
\end{equation}
where $\phi_Q$, $\phi_A$, and $\phi_C$ respectively denote query representation, app representation, and context representation components. $\psi$ is a target apps selection component that takes the mentioned representations and generates an app selection score. These components can be implemented in different ways. In addition, $c_q$ can contain various types of query context, including search time, search location, and the users apps usage.

We implement the component $\phi_Q$ with two major functions: an embedding function $\mathcal{E} : V \rightarrow \mathbb{R}^{d}$ that maps each vocabulary term to a $d$-dimensional embedding space, and a global term weighting function $\mathcal{W}: V \rightarrow \mathbb{R}$ that maps each vocabulary term to a real-valued number showing its global importance. The matrices $\mathcal{E}$ and $\mathcal{W}$ are the network parameters in our model and are learned to provide task-specific representations. The query representation component $\phi_Q$ represents a given query $q = \{w_1, w_2, \cdots, w_{|q|}\}$ as follows:
\begin{equation*}
\phi_Q(q) = \sum_{i=1}^{|q|} \widehat{\mathcal{W}}(w_i)\cdot \mathcal{E}(w_i)~,
\end{equation*}
which is the weighted element-wise summation over the terms' embedding vectors. $\widehat{\mathcal{W}}$ is the normalized global weights computed using a softmax function as follows:
\begin{equation}
\widehat{\mathcal{W}}(w_i) = \frac{\exp(\mathcal{W}(w_i))}{\sum_{j=1}^{|q|}{ \exp(\mathcal{W}(w_j))}}\nonumber~.
\end{equation}

This is a simple yet effective approach for query representation based on the bag of words assumption, which has been proven to be effective for target apps selection \cite{AliannejadiSigir18} and ad-hoc retrieval \cite{Dehghani:2017,DBLP:journals/corr/abs-2009-09392}.

To implement the app representation component $\phi_A$, we learn a $d$-dimensional dense representation for each app. More specifically, this component consists of an app representation matrix $\mathcal{A} \in \mathbb{R}^{N \times d}$ where $N$ denotes the total number of apps. Therefore, $\phi_A(a)$ returns a row of the matrix $\mathcal{A}$ that corresponds to the app $a$. 

Various context definitions can be considered to implement the context representation component. General types of context, such as location and time, has been extensively explored in different tasks, such as web search \cite{DBLP:conf/sigir/BennettRWY11}, personal search \cite{Zamani:2017:WWW}, and mobile search \cite{DBLP:conf/saint/HattoriTT07}. In this paper, we refer to the \emph{apps usage time} as context, which is a special type of context for our task. As introduced earlier in Section \ref{sec:context}, the apps usage context is the time that the user spent on each mobile app in the past 24 hours of the search time. To implement $\phi_C$, we first compute a probabilistic distribution based on the apps usage context, as follows:
\begin{equation*}
    p(a' | c_q) = \frac{\text{time spent on app $a'$ in the past 24 hours}}{\sum_{a'' \in A}{\text{time spent on app $a''$ in the past 24 hours}}}~,
\end{equation*}
where $A$ is a set of candidate apps. $\phi_C$ is then computed as:
\begin{equation*}
    \phi_C(c_q) = \sum_{a' \in A}{p(a' | c_q)\cdot \mathcal{A}_C[a']}~,
\end{equation*}
where $\mathcal{A}_C \in \mathbb{R}^{N \times d}$ denotes an app representation matrix which is different from $\mathcal{A}$ used in the app representation component. This matrix is supposed to learn app representations suitable for representing the apps usage context. $\mathcal{A}_C[a']$ denotes the representation of app $a'$ in the app representation matrix $\mathcal{A}_C$. 

In summary, each of the representation learning components $\phi_Q$, $\phi_A$, and $\phi_C$ returns a $d$-dimensional vector. The app selection component is modeled as a fully-connected feed-forward network with two hidden layers and the output dimensionality of $1$. We use rectified linear unit (ReLU) as the activation function in the hidden layers of the network. Sigmoid is used as the final activation function. To avoid overfitting, the dropout technique~\cite{Srivastava:2014} is employed. For each query, the following vector is fed to this network:
\begin{equation*}
    (\phi_Q(q) \circ \phi_A(a)) \cdot |\phi_Q(q) - \phi_A(a)| \cdot (\phi_C(c_q) \circ \phi_A(a)) \cdot |\phi_C(c_q) - \phi_A(a)|~,
\end{equation*}
where $\circ$ denotes the Hadamard product, i.e., the element-wise multiplication, and $\cdot$ here means concatenation. In fact, this component computes the similarity of the candidate app with the query content and context, and estimates the app selection score based on the combination of both.

We train our model using pointwise and pairwise settings. In a pointwise setting, we use mean squared error (MSE) as the loss function. MSE for a mini-batch $b$ is defined as follows:
\begin{equation}
    \mathcal{L}_{MSE}(b) = \frac{1}{|b|}\sum_{i=1}^{|b|}{(y_i - \psi(\phi_Q(q_i), \phi_A(a_i), \phi_C(c_{q_i})))^2}\nonumber~,
\end{equation}
where $q_i$, $c_{q_i}$, $a_i$, and $y_i$ denote the query, the query context, the candidate app, and the label in the $i$\textsuperscript{th} training instance of the mini-batch. For this training setting, we use a linear activation for the output layer.

\modelname can be also trained in a pairwise fashion. Therefore, each training instance consists of a query, the query context, a target app, and a non-target app. To this end, we employ hinge loss (max-margin loss function) that has been widely used in the learning to rank literature for pairwise models \cite{Li:2011}. Hinge loss is a linear loss function that penalizes examples violating the margin constraint. For a mini-batch $b$, hinge loss is defined as below:
\begin{align}
    \mathcal{L}_{Hinge}(b) = \frac{1}{|b|}\sum_{i=1}^{|b|}&\max \left\{0, 1-\text{sign}(y_{i1} - y_{i2})(\widehat{y}_{i1} - \widehat{y}_{i2} )\right\}\nonumber~,
\end{align}
where $\widehat{y}_{ij} = \psi(\phi_Q(q_i), \phi_A(a_{ij}), \phi_C(c_{q_i}))$.

\begin{figure}
    \centering
    \includegraphics[width=.7\columnwidth]{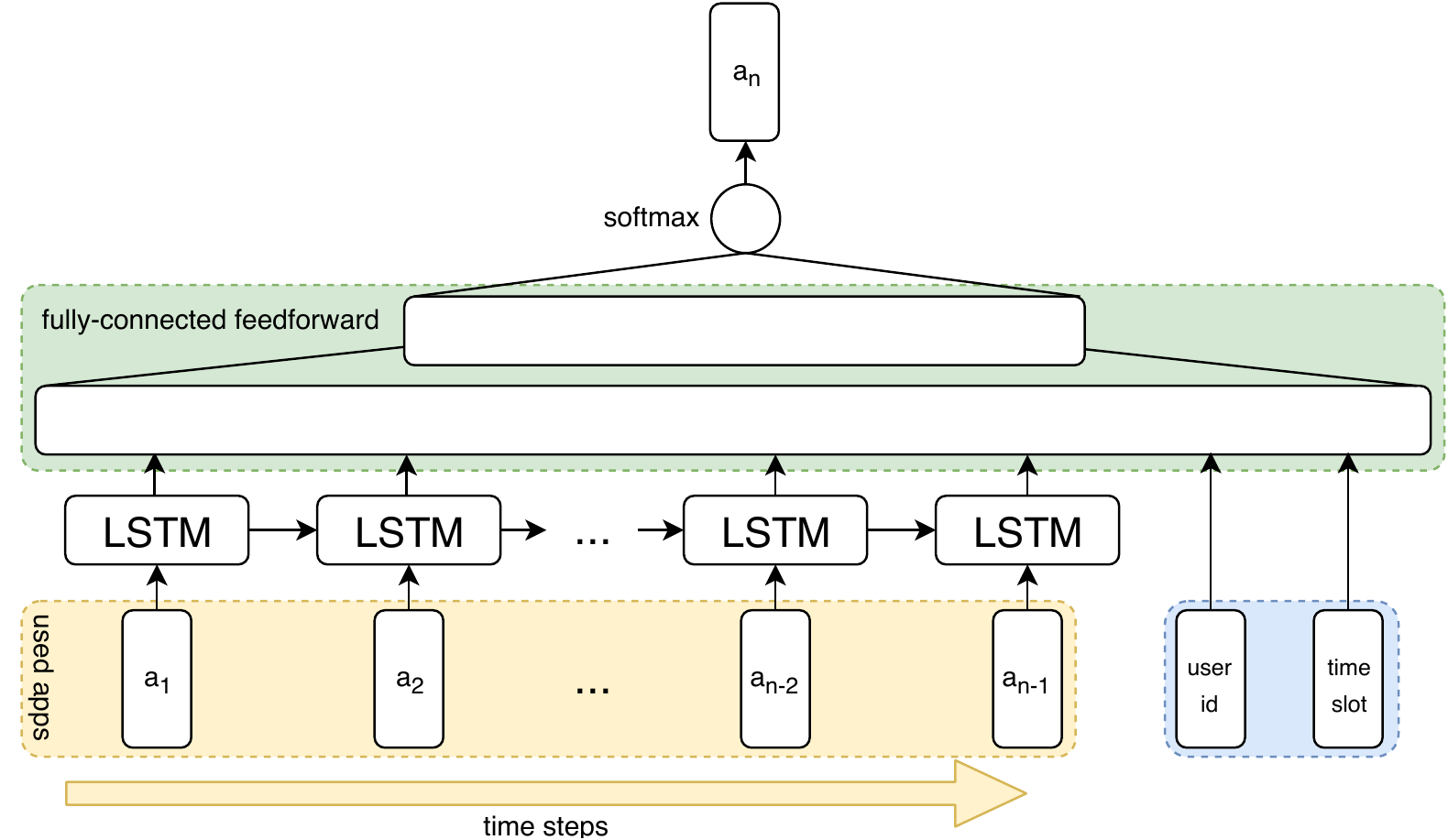}
    \caption{The architecture of \modelnameneusa}
    \label{fig:arch}
\end{figure}

\section{Personalized Time-Aware Target Apps Recommendation}
\label{sec:nextapp}

In this section, we propose a neural sequence-aware model called \modelnameneusa (\textbf{Neu}ral \textbf{S}equential target \textbf{A}pp recommendation), which captures the sequential dependencies of apps as well as users behavior with respect to their usage patterns (i.e., the personal app sequence) and temporal behavior. 
In the following, we first describe an overview of our target apps recommendation and further explain how it is implemented.

Formally, \modelnameneusa estimates the probability $p(L=1 | u, a, c_u; A)$, where $L$ is a binary random variable indicating whether the app $a$ should be launched ($L=1$) or not ($L=0$). $A$ denotes the set of candidate apps. Similar to \modelname, this set can be either all apps, those that are installed on the user's mobile device, or a set of candidate apps that is obtained by another model in a cascade setting. The app recommendation probability in the \modelnameneusa framework is estimated as follows:
\begin{equation*}
    p(L=1 | u, a, c_u; A) = \psi(\phi_U(u), \phi_A(a), \phi_C(c_u))~,
\end{equation*}
where $\phi_U$, $\phi_A$, and $\phi_C(c_u)$ denote user, app, and user context representation components, respectively. $\psi$ is a target apps recommendation component that takes the mentioned representations and generates a recommendation score. Any of these components can be implemented in different ways. In addition, $c_u$ can contain various types of user context, including time, location, and sequence of previously-used apps.

We implement the component $\phi_U$ with an embedding function $\mathcal{E} : \mathcal{U} \rightarrow \mathbb{R}^{d}$ that maps a user  to a $d$-dimensional embedding space. The matrix $\mathcal{E}$ is the network parameter in our model and is learned to provide task-specific representations.

To implement the app representation component $\phi_A$, we learn a $d$-dimensional dense representation for each app. In more detail, this component consists of an app representation matrix $\mathcal{A} \in \mathbb{R}^{N \times d}$ where $N$ denotes the total number of apps. Therefore, $\phi_A(a)$ returns a row of the matrix $\mathcal{A}$ that corresponds to the app $a$. 

General types of context, such as location and time, has been extensively explored in different tasks, such as web search \cite{DBLP:conf/sigir/BennettRWY11}
and mobile search \cite{DBLP:conf/saint/HattoriTT07}. In this paper, we refer to the \emph{$k$ previously-used apps} and time as context with $k=9$.
Therefore, we define a window of size $k$ and consider the sequence of used apps just before the time of recommendation as the sequence context.
Following~\cite{DBLP:journals/corr/BaltrunasCKO15}, we break a full day (i.e., 24 hours) into 8 equal time bins (early morning - late night). 
To implement $\phi_C$, we first compute a probabilistic distribution based on the apps usage records, as follows:
\begin{equation*}
    p(a' | c_u) = \frac{\text{time spent on app $a'$ in the current time bin}}{\sum_{a'' \in A}{\text{time spent on app $a''$ in the current time bin}}}~,
\end{equation*}
where $A$ is a set of candidate apps. $\phi_C$ is then computed as:
\begin{equation*}
    \phi_C(c_u) = \sum_{a' \in A}{p(a' | c_u)\cdot \mathcal{A}[a']}~,
\end{equation*}
where $\mathcal{A} \in \mathbb{R}^{N \times d}$ denotes an app representation matrix. This matrix is supposed to learn app representations suitable for representing sequences of apps. $\mathcal{A}[a']$ denotes the representation of app $a'$ in the app representation matrix $\mathcal{A}$. 
 
Each of the representation learning components $\phi_U$, $\phi_A$, and $\phi_C$ returns a $d$-dimensional vector. The app recommendation component is modeled as a recurrent neural network (RNN) consisting of Long Short-Term Memory (LSTM) units. After modeling the sequence of apps in this layer, the parameters, together with user and time features are passed to a fully-connected feedforward network with two hidden layers. We use rectified linear unit (ReLU) as the activation function in the hidden layers of the network. Softmax is used as the final activation function. To avoid overfitting, the dropout technique~\cite{Srivastava:2014} is employed.
We train our model using pointwise training setting where we use cross entropy as the loss function.
Figure~\ref{fig:arch} depicts the architecture of our proposed network.

\section{Experimental Setup}
\label{sec:expsetup}

In this section, we evaluate the performance of the proposed model in comparison with a set of baseline models.
\subsection{Target Apps Selection}

\partitle{Data.}
We evaluate the performance of our proposed models on the \istas~dataset. We follow two different strategies to split the data:
\begin{inlinelist}
    \item In \textit{\istas-R}, we randomly select 70\% of the queries for training, 10\% for validation, and 20\% for testing;
    \item In \textit{\istas-T}, we sort chronologically the queries of each user and keep the first 70\% of each user's queries for training, the next 10\% for validation, and the last 20\% for testing. \istas-T is used to evaluate the methods when information about users' search history is available.
\end{inlinelist} 
To minimize random bias, for \istas-R we repeated the experiments 10 times and report the average performance. 
The hyper-parameters of all models were tuned based on the nDCG@3 value on the validation sets.

\partitle{Evaluation metrics.}
Effectiveness is measured by four standard evaluation metrics that were also used in~\cite{AliannejadiSigir18}: mean reciprocal rank (MRR), and normalized discounted cumulative gain for the top 1, 3, and 5 retrieved apps (nDCG@1, nDCG@3, nDCG@5). We determine the statistically significant differences using the two-tailed paired t-test with Bonferroni correction at a $95\%$ confidence interval ($p < 0.05$). 

\partitle{Compared methods.} 
We compared the performance of our model with the following methods:
\begin{itemize}[leftmargin=*]
    \item \textit{MFU (Most Frequently Used):} For every query we rank the apps in the order of their popularity in the training set as a static (query independent) model.
    \item \textit{QueryLM, BM25, BM25-QE:} For every app we aggregate all the relevant queries from the training set to build a document representing the app. QueryLM is the query likelihood retrieval model~\cite{DBLP:conf/sigir/PonteC98}. For BM25-QE, we adopt Bo1~\cite{DBLP:phd/ethos/Amati03} for query expansion. We use the Terrier \cite{DBLP:conf/ecir/OunisAPHMJ05} implementation of these methods.
    \item \textit{k-NN, k-NN-AWE:} To find the nearest neighbors in k nearest neighbors (k-NN), we consider the cosine similarity between the TF-IDF vectors of queries. Then, we take the labels (apps) of the nearest queries and produce the app ranking. As for k-NN-AWE \cite{Zamani:2016}, we compute the cosine similarity between the average word embedding (AWE) of the queries obtained from GloVe~\cite{pennington2014glove} with 300 dimensions.
    \item \textit{ListNet, ListNet-CX:} For every query-app pair, we use the scores obtained by BM25-QE, k-NN, k-NN-AWE, and MFU as features to train ListNet~\cite{DBLP:conf/icml/CaoQLTL07} implemented in RankLib\footnote{\url{https://sourceforge.net/p/lemur/wiki/RankLib/}}. For every query, we consider all irrelevant apps as negative samples. ListNet-CX also includes users' apps usage context as an additional feature.
    \item \textit{\ntar}: A neural model approach that we designed for the target apps selection task in our previous work~\cite{AliannejadiSigir18}. We use the NTAS1 model due to its superior performance compared to NTAS2.
    \item \textit{Contextual baselines:} In order to carry out a fair comparison between \modelname and other context-aware baselines, we apply a context filter to all non-contextual baselines. We create the context filter as follows: for every app $\alpha$ in the training samples of user $u$, we take the time that $u$ has spent on $\alpha$ in the past 24 hours as its score. We then perform a linear interpolation with the scores of all the mentioned baselines. Note that all scores are normalized. All these models are denoted by a \textit{-CR} suffix.
\end{itemize}

\subsection{Target Apps Recommendation}

\partitle{Data.}
For every user, we take the 70\% earliest app usage records as training set, 10\% next records as validation, and 20\% latest records as test set.

\partitle{Evaluation metrics.}
Effectiveness is measured by 6 standard evaluation metrics: mean reciprocal rank (MRR), normalized discounted cumulative gain for the top 1, 3, and 5 predicted apps (nDCG@1, nDCG@3, nDCG@5), and recall for the top 3 and 5 predicted apps (Recall@3, Recall@5). 
Our choice of evaluation metrics was motivated by the two main purposes of app recommendation we discussed in Section~\ref{sec:intro}. The MRR and nDCG@$k$ metrics are intended to evaluate the effectiveness for improved homescreen app ranking user experience, whereas Recall@$k$ mainly evaluates how well a model is able to pre-load the next app among the top $k$ predicted apps. 

We determine the statistically significant differences using the two-tailed paired t-test at a 99.9\% confidence interval ($p < 0.001$). Note that we apply the Bonferroni correction for the test against the baselines (i.e., * in Table~\ref{tab:rec-results}).

We compare the performance of our models with the following methods:
\begin{itemize}[leftmargin=5pt]
    \item \textit{MFU (Most Frequently Used)}: For every test instance we rank the apps in the order of their popularity in the training set as a static recommendation model.
    \item \textit{MRU (Most Recently Used)}: For every test instance we rank the apps in the order of the their interaction time, so that the most recent apps are ranked higher.
    \item \textit{\revision{Bayesian \& Linear \cite{DBLP:conf/huc/HuangZMC12}}}: \revision{We implement the two baselines proposed by \citet{DBLP:conf/huc/HuangZMC12}, namely, Bayesian and Linear. Both baselines incorporate various contextual information in modeling app usage. In this work, we only use the contextual information available in our dataset, i.e., time, weekday, user, and previous app.}
    \item \textit{LambdaMART \& ListNet}: For a given candidate app and every app in the sequence context, we compute the cosine similarity of their representation and consider it as a feature. The app representations are the average word embedding (AWE) of app descriptions on Google Play Store. Other features include the recommendation time and current user. These features were used to train LambdaMART and ListNet as state-of-the-art learning to rank (LTR) methods, implemented in RankLib.\footnote{\url{https://sourceforge.net/lemur/wiki/RankLib/}}
    \item \textit{k-NN \& DecisionTree}: Similar to LTR baselines, we take AWE similarity between app pairs as well as user and time as classification features. We also include the apps that appear in the context sequence as additional features. We train kNN and DecisionTree classifiers implemented in scikit-learn.\footnote{\url{https://scikit-learn.org/}}
    \item TempoLSTM~\cite{DBLP:conf/wasa/XuLZGZZZS18} models the sequence of apps using a two-layer network of LSTM units. The temporal information as well as the application is directly passed to each LSTM node.
    \item \textit{\modelnameneusa$_{\text{w/o user}}$, \modelnameneusa$_{\text{w/o time}}$ \& \modelnameneusa$_{\text{w/o user, w/o time}}$}: These are three variations of our model. 
    The only difference is in the use of time and user features in the models. \modelnameneusa$_{\text{w/o user}}$ is trained without user data; \modelnameneusa$_{\text{w/o time}}$ without time data; and \modelnameneusa$_{\text{w/o user, w/o time}}$ without neither of them.
\end{itemize}

\begin{table*}[t]
    \centering
    \caption{Performance comparison with baselines on \istas-R and \istas-T datasets. The superscript * denotes significant differences compared to all the baselines.}
    \label{tab:sel-results}
    \resizebox{13.85cm}{!}{%
    \begin{tabular}{lccccccccc}
    \toprule
     \multirow{2}{*}{\textbf{Methods}} & \multicolumn{4}{c}{\textbf{\istas-R Dataset}} && \multicolumn{4}{c}{\textbf{\istas-T Dataset}} \\
     \cmidrule{2-5} \cmidrule{7-10} 
     & MRR & nDCG@1 & nDCG@3 & nDCG@5 && MRR & nDCG@1 & nDCG@3 & nDCG@5\\
    \midrule \midrule
    MFU & 0.4502 & 0.2597 & 0.4435 & 0.4891 && 0.4786 & 0.2884 & 0.4752 & 0.5173 \\
    
    QueryLM & 0.3556 & 0.2431 & 0.3534 & 0.3900 && 0.2706 & 0.1486 & 0.2713 & 0.3097 \\
    BM25 & 0.4205 & 0.3134 & 0.4363 & 0.4564 && 0.3573 & 0.2447 & 0.3771 & 0.3948 \\
    BM25-QE & 0.4319 & 0.2857 & 0.4371 & 0.4727 && 0.3930 & 0.2411 & 0.4053 & 0.4364 \\
    k-NN & 0.4433 & 0.2761 & 0.4455 & 0.4811 && 0.4067 & 0.2294 & 0.3982 & 0.4655 \\
    k-NN-AWE & 0.4742 & 0.2937 & 0.4815 & 0.5211 && 0.4859 & 0.2950 & 0.4919 & 0.5392 \\
    ListNet & 0.5170 & 0.3330 & 0.5211 & \textbf{0.5623} && 0.5118 & 0.3219 & \textbf{0.5208} & 0.5572 \\
    NTAS-pointwise & 0.5221 & 0.3427 & 0.5231 & 0.5586 && 0.5162 & 0.3385 & 0.5162 & 0.5550 \\
    NTAS-pairwise & \textbf{0.5257} & \textbf{0.3468} & \textbf{0.5236} & 0.5618 && \textbf{0.5214} & \textbf{0.3427} & 0.5183 & \textbf{0.5580} \\
    \midrule \midrule
    \textbf{Context-Aware Methods} \\
    \midrule
    MFU-CR & 0.4903 & 0.3015 & 0.4901 & 0.5268 && 0.5289 & 0.3576 & 0.5358 & 0.5573 \\
    QueryLM-CR & 0.4540 & 0.2773 & 0.4426 & 0.5013 && 0.4696 & 0.3023 & 0.4597 & 0.5145 \\
    BM25-CR & 0.5398 & 0.3653 & 0.5394 & 0.5871 && 0.5249 & 0.3496 & 0.5255 & 0.5723 \\
    BM25-QE-CR & 0.5215 & 0.3398 & 0.5223 & 0.5693 && 0.5230 & 0.3474 & 0.5260 & 0.5728 \\
    k-NN-CR & 0.4978 & 0.3114 & 0.4926 & 0.5431 && 0.5161 & 0.3481 & 0.4956 & 0.5555 \\
    k-NN-AWE-CR & 0.5144 & 0.3233 & 0.5142 & 0.5632 && 0.5577 & 0.3722 & 0.5612 & \textbf{0.6086} \\
    ListNet-CR & 0.5391 & 0.3544 & 0.5417 & 0.5845 && 0.5599 & 0.3780 & 0.5657 & 0.6037 \\
    ListNet-CX & 0.5349 & 0.3580 & 0.5343 & 0.5784 && 0.5019 & 0.3139 & 0.5153 & 0.5521 \\
    NTAS-pointwise-CR & 0.5532 & 0.3745 & 0.5580 & 0.5883 && 0.5627 & 0.3865 & 0.5663 & 0.5965 \\
    NTAS-pairwise-CR & 0.5576 & 0.3779 & 0.5568 & 0.5870 && 0.5683 & 0.3923 & 0.5661 & 0.6047 \\
    \modelname-pointwise & 0.5614* & 0.3833* & \textbf{0.5592} & 0.5901 && 0.5702* & 0.4146* & 0.5655 & 0.5938 \\
    \modelname-pairwise & \textbf{0.5637}* & \textbf{0.3861}* & 0.5586 & \textbf{0.5924}* && \textbf{0.5738}* & \textbf{0.4182}* & \textbf{0.5679} & 0.6071 \\
    \bottomrule
    \end{tabular}
    }
\end{table*}

\section{Results and Discussion}
\label{sec:res}

In the following, we evaluate the performance of \modelname trained on both data splits and study the impact of context on the performance. We further analyze how the models perform on both data splits.

\subsection{Target Apps Selection}

\partitle{Performance comparison.}
Table~\ref{tab:sel-results} lists the performance of our proposed methods versus the compared methods. 
First, we compare the relative performance drop between the two data splits.
We see that almost all non-contextual models perform worse on \istas-T compared to \istas-R, whereas almost all context-aware models perform better on \istas-T. Among the non-contextual methods, ListNet is the most robust model with the lowest performance drop and k-NN-AWE is the only method that performs better on \istas-T (apart from MFU). Worse results achieved by MFU suggests that \istas-T is less biased towards most popular apps, hence being more challenging. On the other hand, QueryLM exhibits the highest performance drop ($-27\%$ on average), as opposed to Contextual-k-NN-AWE with the highest performance improvement on \istas-T ($+$10\% on average). This indicates that k-NN-AWE is able to capture similar queries effectively, whereas QueryLM relies heavily on the indexed queries. It should also be noted that MFU performs better on \istas-T indicating that it is more biased towards popular apps.

Among the non-contextual baselines, we see that \ntar-pairwise performs best in terms of most evaluation metrics on both data splits, this is because it learns high dimensional app and query representations which help it to perform more effectively.
We see that applying the contextual filter improves the performance of all models. These improvements are statistically significant in all cases, so are not shown in the table. Although this filter is very simple, it is still able to incorporate useful information about user context and behavior into the ranking. This also indicates the importance of apps usage context, as mentioned in Section \ref{sec:context}.
Among the context-aware baselines, we see that \ntar-pairwise-CR performs best in terms of MRR and nDCG@1, while k-NN-AWE-CR and ListNet-CR perform better in terms of other evaluation metrics. It should also be noted that ListNet-CR performs better than ListNet-CX. This happens due to the fact that ListNet-CX integrates the apps usage context as an additional feature, whereas ListNet-CR is the result of the combination of ListNet and the contextual filter. 

\begin{figure}
    \centering
    \includegraphics[width=.8\textwidth,valign=t]{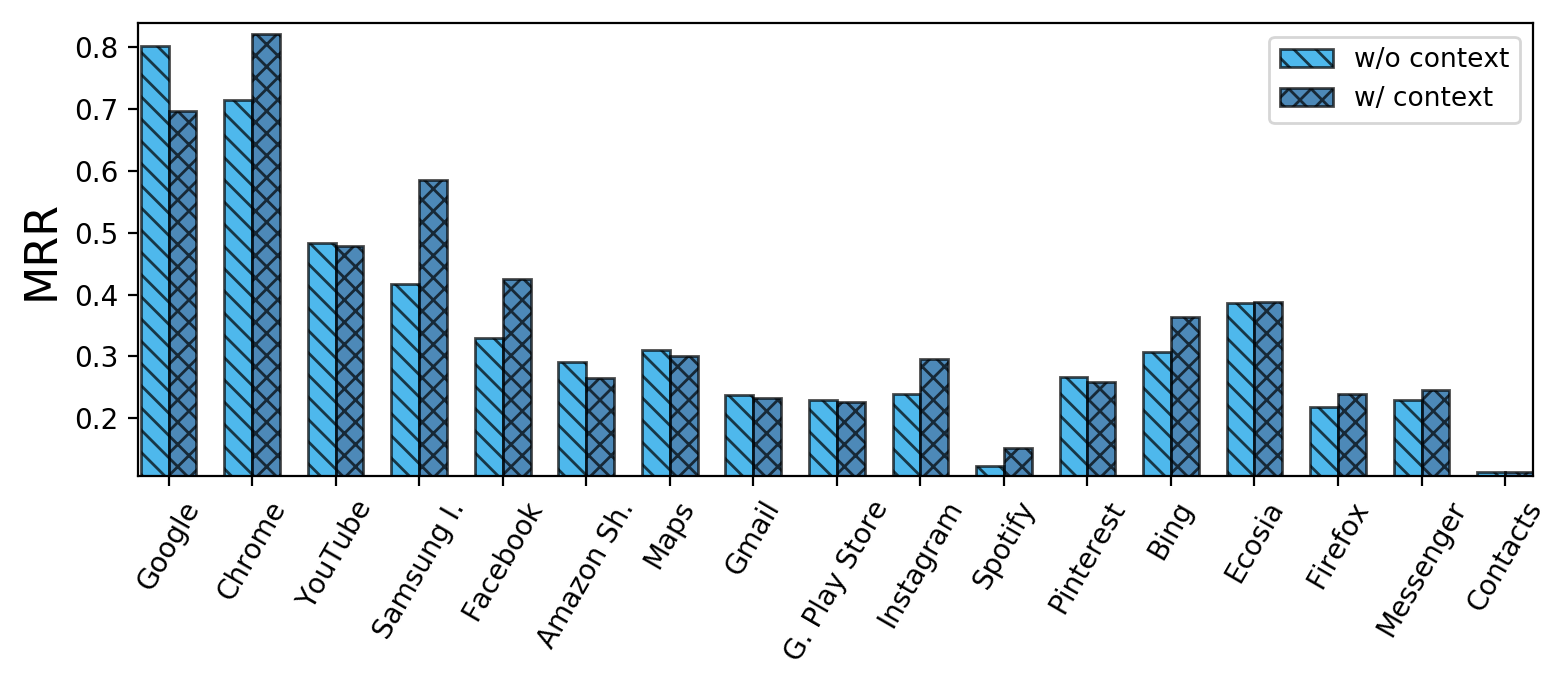}
    \caption{Performance comparison with respect to certain apps with and without context.}
    \label{fig:res_app}
\end{figure}

We see that our proposed \modelname outperforms all the baselines with respect to the majority of evaluation metrics. In particular \modelname-pairwise exhibits the best performance. The achieved improvements in terms of MRR and nDCG@1 are statistically significant. The reason is that \modelname is able to learn latent features from the interaction of mobile usage data in the context. These interactions can reveal better information for better understanding the user information needs. 

\begin{figure}
    \centering
    \subfloat[$\Delta$MRR per app]{
        \includegraphics[height=0.2\textheight,valign=t]{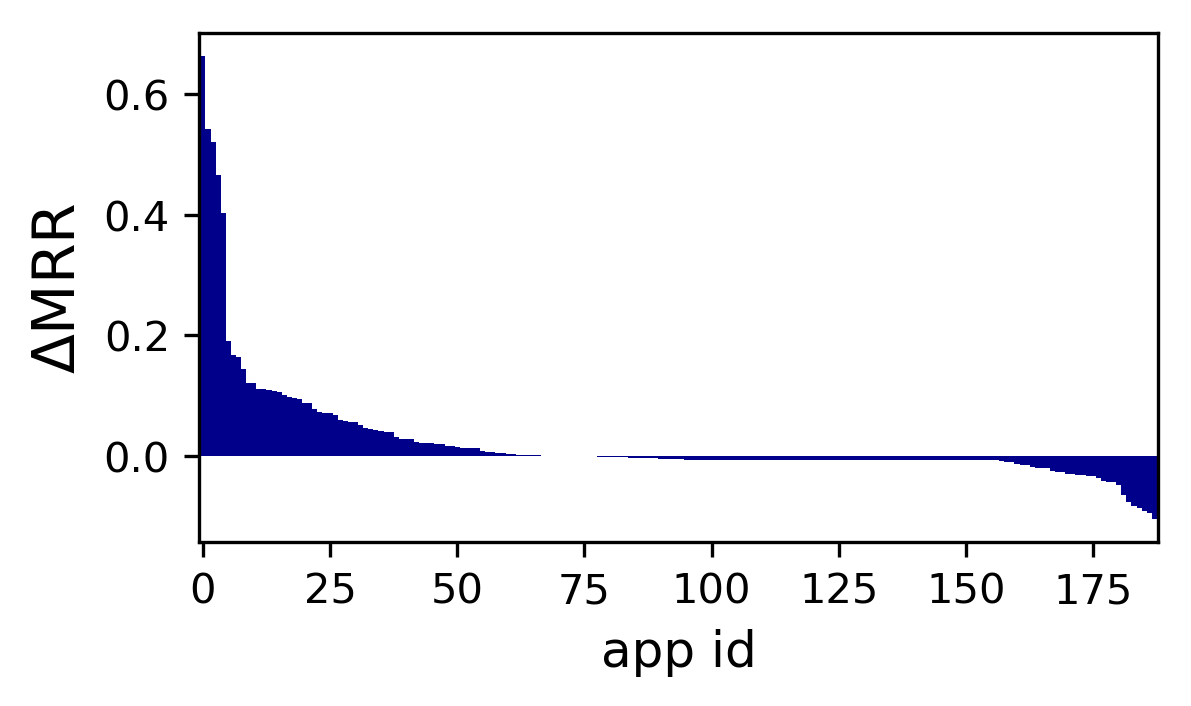}
        \label{fig:diff_app}
    }
    \subfloat[$\Delta$MRR per user]{
        \includegraphics[height=0.2\textheight,valign=t]{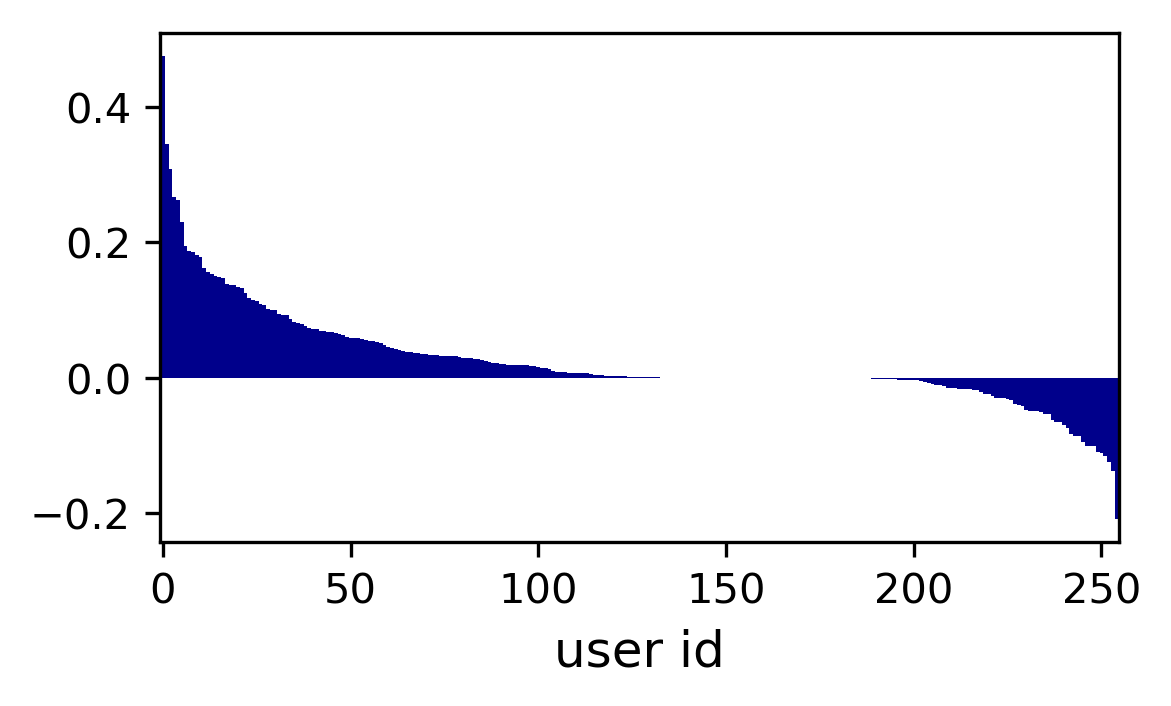}
        \label{fig:diff_user}
    }
    \caption{Performance difference per app and per user in terms of $\Delta$MRR.}
    \label{fig:res_task}
\end{figure}

\partitle{Impact of context on performance per app.}
In this experiment we demonstrate the effect of context on the performance with respect to various apps. Figure~\ref{fig:res_app} shows the performance for queries that are labeled for specific target apps (as listed in the figure). We see that the context-aware model performs better while predicting social media apps such as \appname{Facebook} and \appname{Instagram}. However, we see that the performance for \appname{Google} drops as it improves for \appname{Chrome}. This happens because users do most of their browsing activities on \appname{Chrome}, rather than on  \appname{Google}; hence the usage statistics of \appname{Chrome} helps the model to predict it more effectively. Moreover, we study the difference of MRR between the model with and without context for all apps. Our goal is to see how context improves the performance for every target app. We see in Figure \ref{fig:diff_app} that the performance is improved for 39\% of the apps. As shown in the figure, the improvements are much larger compared with the performance drops. Among the apps with the highest context improvements, we can mention \appname{Quora}, \appname{Periscope}, and \appname{Inbox}. 

\partitle{Impact of context on performance per user.} 
Here we study the difference of MRR between the model with and without context for all users. Our goal is to see how many users are impacted positively by incorporating context in the target apps selection model. Figure \ref{fig:diff_user} shows how performance differs per user when we apply context compared with when we do not. As we can see, users' apps usage context is able to improve the effectiveness of target apps selection for the majority of users. In particular, the performance for 57\% of the users is improved by incorporating the apps usage context. In fact, we observed that users with the highest impact from context use less popular apps.

\begin{table}[t]
    \centering
    \caption{Performance analysis based on query length, dividing the test queries into three evenly-sized length buckets.}
    \begin{tabular}{llll}
        \toprule
         & \multicolumn{1}{c}{Short queries} &  \multicolumn{1}{c}{Med. queries} &  \multicolumn{1}{c}{Long queries} \\
         \cmidrule{2-4}
         & MRR & MRR & MRR \\
         \midrule
         w/o context & 0.5302 & 0.4924 & 0.4971 \\
         w/ context & \textbf{0.5733} & \textbf{0.5190} & \textbf{0.4977}\\
         \bottomrule
    \end{tabular}
    \label{tab:query_len}
\end{table}

\partitle{Impact of context on performance per query length.}
We create three buckets of test queries based on query length uniformly. Therefore, the buckets will have approximately equal number of queries. The first bucket, called Short queries, contains the shortest queries, the second one, called Med. queries, constitutes of medium-length queries and the last bucket, called Long queries, obviously includes the longest queries of our test set. Table \ref{tab:query_len} lists the performance of the model with and without context in terms of MRR. As we can see, the average MRR for all three buckets is improved as we apply context. However, we observe that as the queries become shorter, the improvement increases. The reason is that shorter queries tend to be more general or ambiguous, and thus query context can have higher impact on improving search for these queries.

\begin{table*}[]
    \centering
    \caption{Performance comparison with baselines on the \lsapp dataset. The superscripts $\ast$, $\star$, $\dagger$, and $\ddagger$ denote significant improvements compared to all the baselines, \modelnameneusa$_{\text{w/o user, w/o time}}$, \modelnameneusa$_{\text{w/o user}}$, and \modelnameneusa$_{\text{w/o time}}$ respectively ($p < 0.001$).}
    \label{tab:rec-results}
    \resizebox{13.9cm}{!}{%
    \begin{tabular}{lllllll}
    \toprule
     \textbf{Method} & \textbf{MRR} & \textbf{nDCG@1} & \textbf{nDCG@3} & \textbf{nDCG@5}  & \textbf{Recall@3} & \textbf{Recall@5} \\
    \midrule
    \textbf{MFU} & 0.2630 & 0.1009 & 0.2800 & 0.3413 & 0.3229 & 0.4554 \\
    \textbf{MRU} & 0.3652 & 0.0323 & 0.3232 & 0.4106 & 0.2928 & 0.4898 \\
    \revision{\textbf{Bayesian~\cite{DBLP:conf/huc/HuangZMC12}}} & 0.1461 & 0.0679 & 0.1506 & 0.1687 & 0.1664 & 0.2047 \\
    \revision{\textbf{Linear~\cite{DBLP:conf/huc/HuangZMC12}}} & 0.1363 & 0.0592 & 0.1433 & 0.1626 & 0.1636 & 0.2043 \\
    \textbf{LambdaMART} & 0.3809 & 0.2257 & 0.4532 & 0.4739 & 0.4734 & 0.4958 \\
    \textbf{ListNet} & 0.4992 & 0.3908 & 0.5477 & 0.5683 &  0.5632 & 0.6069 \\
    
    \textbf{k-NN} & 0.4824 & 0.3699 & 0.5158 & 0.5519 & 0.5396 & 0.6165 \\
    \textbf{DecisionTree} & 0.5025 & 0.4422 & 0.5315 & 0.5375 & 0.5343 & 0.5471 \\
    \textbf{TempoLSTM~\cite{DBLP:conf/wasa/XuLZGZZZS18}} & 0.6869 & 0.5715 & 0.7424 & 0.7730 & 0.7668 & 0.8316 \\
    \midrule
    \textbf{\modelnameneusa}$_{\text{w/o user, w/o time}}$   & 0.6924$^\ast$ & 0.5677 & 0.7539$^\ast$ & 0.7817$^\ast$ & 0.7817$^\ast$ & 0.8547$^\ast$ \\
    \textbf{\modelnameneusa}$_{\text{w/o user}}$ & 0.6971$^{\ast\star}$ & 0.5721$^{\star}$ & 0.7592$^{\ast\star}$ & 0.7940$^{\ast\star}$ & 0.7873$^{\ast\star}$ & 0.8608$^{\ast\star}$ \\
    \textbf{\modelnameneusa}$_{\text{w/o time}}$ & 0.7036$^{\ast\star\dagger}$ & 0.5745$^*$ & 0.7672$^{\ast\star\dagger}$ & 0.8053$^{\ast\star\dagger}$ & 0.7964$^{\ast\star\dagger}$ & 0.8770$^{\ast\star\dagger}$ \\
    \textbf{\modelnameneusa} & 0.7049$^{\ast\star\dagger\ddagger}$ & 0.5763$^{\ast\star\dagger}$ & 0.7680$^{\ast\star\dagger}$ & 0.8062$^{\ast\star\dagger}$ & 0.7969$^{\ast\star\dagger}$ & 0.8779$^{\ast\star\dagger}$ \\
    \bottomrule
    \end{tabular}
    }
    \vspace{-0.3cm}
\end{table*}

\subsection{Target Apps Recommendation}
In the following, we evaluate the performance of \modelnameneusa trained on \lsapp and study the impact of time and user features as well as of the learned app representations.

\partitle{Performance comparison.}
Table~\ref{tab:rec-results} lists the performance of our proposed method as well as its variations and baselines. As we can see, ListNet exhibits the best performance among LTR baselines and DecisionTree among classification baselines. Moreover, all models outperform MFU in terms of all evaluation metrics. In particular, we see that Recall@5 is improved for all methods, indicating that allowing most used apps to run in the background is not effective. Also, we see that while ListNet consistently outperforms LambdaMART, k-NN exhibits a better performance than DecisionTree in terms on Recall@3 and Recall@5. \revision{We see that all models, including MFU and MRU, outperform the statistical baselines, namely, Bayesian and Linear. The large margin in the performance of simple models such as k-NN with these two models indicates the effectiveness of representation-based features (i.e., AWE similarity) for this task.} Furthermore, we see that \modelnameneusa outperforms all the baselines by a large margin in terms of all evaluation metrics.
For instance, we see a 39\% relative improvement over DecisionTree in terms of MRR and a 40\% relative improvement over k-NN in terms of Recall@5. This suggests that learning high dimensional sequence-aware representation of apps enables the model to capture users behavioral patterns in using their smartphones. It is worth noting that \modelnameneusa achieves a high value of Recall@5, suggesting that a mobile operating system is able to pre-load 5 apps with an 87\% recall value.

\partitle{Impact of time and user features.}
To evaluate the impact of time and user features we compare the performance of \modelnameneusa with three variations called \modelnameneusa$_{\text{w/o user}}$, \modelnameneusa$_{\text{w/o time}}$, and \modelnameneusa$_{\text{w/o user, w/o time}}$. As we described earlier, these three models are trained after removing user and time features from the data. We see that in all cases, the performance consistently drops. In particular, we see that when both user and time features are removed, \modelnameneusa$_{\text{w/o user, w/o time}}$ exhibits the largest performance loss, while still outperforming all the baseline models for the majority of metrics. As we add the user feature to the model, we see that the performance improves, showing that a personalized app recommendation model is effective. 
In particular, we see that \modelnameneusa$_{\text{w/o time}}$ outperforms \modelnameneusa$_{\text{w/o user, w/o time}}$ significantly in terms of all evaluation metrics. Also, we see a large drop of performance when we remove the user data from \modelnameneusa, confirming again that personal app usage patterns should be taken into consideration for this problem. 
Therefore, a practical system can be trained on a large dataset of app usage from various users and be fine-tuned on every user's phone according to their personal usage behavior.
Furthermore, although we see that adding time to \modelnameneusa$_{\text{w/o user, w/o time}}$ model results in significant improvements (i.e., \modelnameneusa$_{\text{w/o user}}$), we do not observe the same impact after adding the user data to the model (comparing \modelnameneusa against \modelnameneusa$_{\text{w/o time}}$). This suggests that while temporal information contain important information revealing time-dependent app usage patterns, it does not add useful information to the personal model. This can be due to the fact that the personal information already conveys the temporal app usage behavior of the user (i.e., each user temporal behavior is unique).

\begin{figure}
    \centering
    \includegraphics[width=0.5\linewidth]{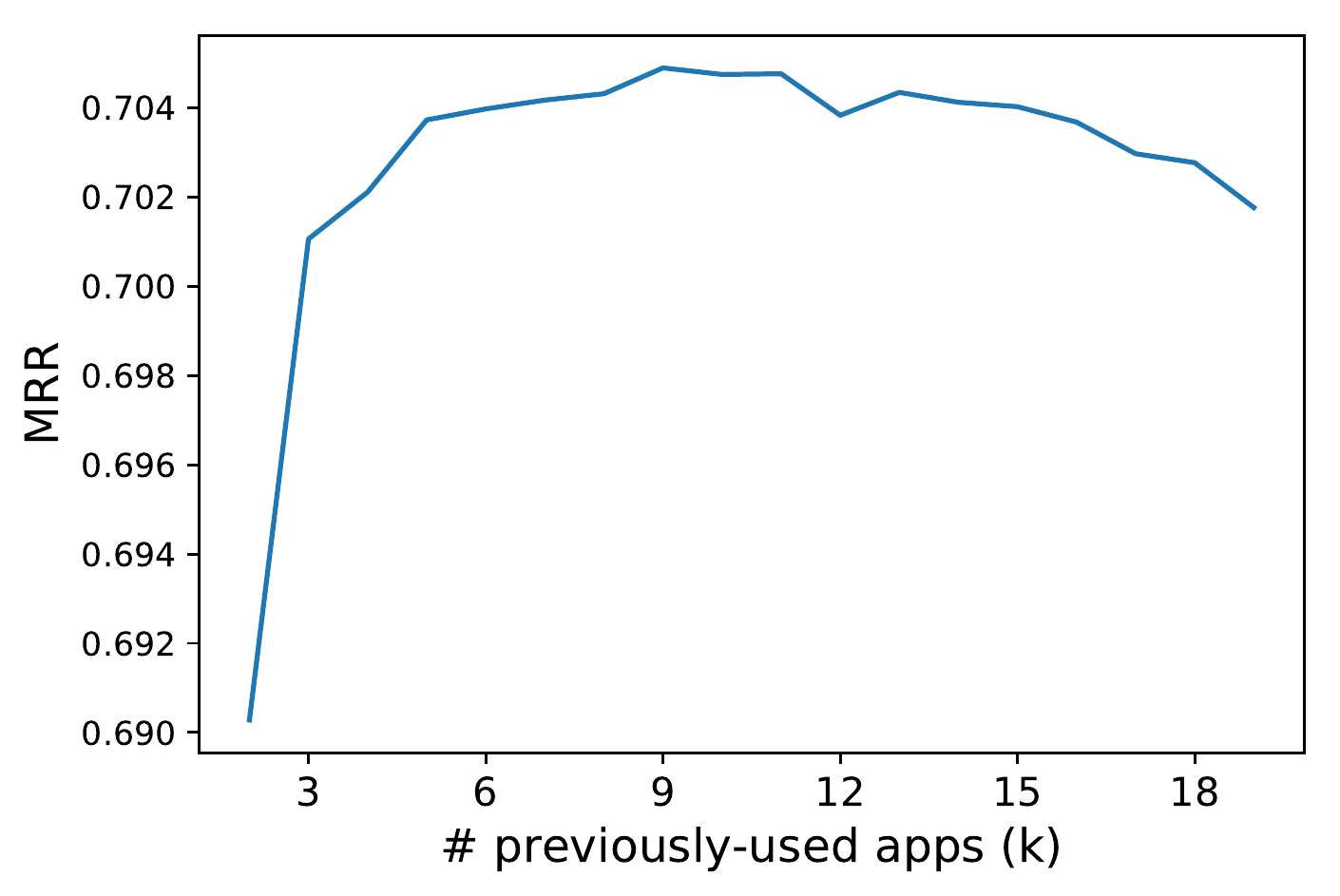}
    \caption{Performance of \modelnameneusa in terms of MRR for different number of previously-used apps as context ($k$).}
    \label{fig:hist_len}
\end{figure}

\partitle{Impact of number of the context length.}
Here, we evaluate the effect of the number of previously-used apps that we consider in our \modelnameneusa model. To do so, we keep all the model parameters the same and change the number of apps in the context ($k$). We plot the performance of \modelnameneusa for various $k$ values in Figure~\ref{fig:hist_len}. As we see in the figure, even though the performance somewhat converges with $k \geq 3$, the best performance is achieved with $k=9$. This indicates that while the model depends highly on the latest three apps that have been used by the user, it can learn some longer patterns in some rare cases. Moreover, it is worth noting that the model's performance using only one app in the context in terms of MRR is $0.5509$, indicating that using only one app is not enough for accurate prediction of next-app usage.

\section{Conclusions and Future Work}
\label{sec:conclusion}

In this paper, we conducted the first in situ study on the task of target apps selection, which was motivated by the growing interest in intelligent assistants and conversational search systems where users interact with a universal voice-based search system~\cite{DBLP:conf/sigir/AliannejadiZCC19,DBLP:conf/chiir/AliannejadiCRC20,Macaw,DBLP:conf/ictir/KrasakisAVK20,DBLP:journals/corr/abs-2009-11352}. 
To this aim, we developed an app, \app, and recruited 255 participants, asking them to report their real-life cross-app mobile queries via \app.
We observed notable differences in length and structure among queries submitted to different apps. Furthermore, we found that while users search using various apps, a few apps attract most of the search queries. We found that   even though \appname{Google} and \appname{Chrome} are the most popular apps, users do only 26\% and 23\% of their searches in these apps, respectively. The in situ data collection enabled us to collect valuable information about users' contexts. For instance, we found that the target app for 29\% of the queries were among the top two most used apps of a particular user. Inspired by our data analysis, we proposed a model that learns high-dimensional latent representations for the apps usage context and predicts the target app for a query. The model was trained with an end-to-end setting. Our model produces a score for a given context-query-app triple. We compared the performance of our proposed method with state-of-the-art retrieval baselines splitting data following two different strategies. We observed that our approach outperforms all baselines, significantly. 

Furthermore, we proposed a neural sequence-aware model, called \modelnameneusa, for predicting next app usage. \modelnameneusa learns a high-dimensional representation for mobile apps, incorporating the app usage sequence as well as temporal and personal information into the model. We trained the model on the app usage data collected from 292 real users. The results showed that the proposed model is able to capture complex user behavioral patterns while using their phones, outperforming classification and LTR baselines significantly in terms of nDCG@$k$, Recall@$k$, and MRR.

\partitle{Limitations.}
Like any other study, our study has some limitations. First, the study relies on self-reporting. This could result in specific biases in the collected data. For instance, participants may prefer to report shorter queries simply because it requires less work. Also, in many cases, participants are likely to forget reporting queries or do not report all the queries that belong to the same session. Second, the reported queries are not actually submitted to a unified search system and users may formulate their queries differently is such setting. For example, in a unified system a query may be ``videos of Joe Bonamassa'' but in \appname{YouTube} it may be ``Joe Bonamassa.'' 
Both the mentioned limitations are mainly due to lack of an existing unified mobile search app. Hence, building such app would be essential for building a more realistic collection.
Also, our study does not consider the users' success or failure in their search. Submitting queries in certain apps could result in different chances of success, and consequently, affect users' behavior in the session to submit other queries in the same app or other apps. 
 Finally, more efficient data collection strategies could be employed based on active learning~\cite{DBLP:conf/www/RahmanKL19}.

\partitle{Future work.}
The next step in this research would be exploring the influence of other types of contextual information, such as location and time, on the target apps selection and recommendation tasks. In addition, it would be interesting to explore result aggregation and presentation in the future, considering two important factors: information gain and user satisfaction. This direction can be studied in both areas of information retrieval and human-computer interaction. 
Furthermore, based on our findings in the analyses, we believe that mobile search queries can be leveraged to improve the user experience. For instance, assuming a user searches for a restaurant using a unified search system and finds some relevant information on \appname{Yelp}. In this case, considering the user's personal preference as well as the context, the system could send the user a notification with information about the traffic near the restaurant. This would certainly improve the quality of the user experience. 
We also plan to investigate if the demographics of the participants are linked to particular queries and behavior. And if such behavioral biases exist, how different models are able to address such issues?

\partitle{Acknowledgements.} 
We thank the anonymous reviewers for the valuable feedback. This work was supported in part by the RelMobIR project of the \grantsponsor{}{Swiss National Science Foundation (SNSF)}{http://www.snf.ch/en/Pages/default.aspx}, and in part by the Center for Intelligent Information Retrieval. Any opinions, findings and conclusions or recommendations expressed in this material are those of the authors and do not necessarily reflect those of the sponsors.

\bibliographystyle{ACM-Reference-Format}
\bibliography{sigproc} 

\end{document}